\documentclass{aa}
\bibpunct{(}{)}{;}{a}{}{,} 

\usepackage{graphicx}
\usepackage[varg]{txfonts}
\usepackage[dvipsnames]{xcolor}
\usepackage{caption}

\begin{document} 
\title{An Ingot-like class of WaveFront Sensors for Laser Guide Stars}
\author{
R. Ragazzoni\inst{1,2,6}
\and E. Portaluri\inst{3,6} 
\and D. Greggio\inst{2,6}
\and M. Dima\inst{2,6} 
\and C. Arcidiacono\inst{2,6}
\and M. Bergomi\inst{2,6} 
\and S. Di Filippo \inst{2,6} 
\and T. S. Gomes Machado\inst{1,2,6}
\and K. K. R. Santhakumari\inst{2,6}
\and V. Viotto\inst{2,6}
\and F. Battaini\inst{2,6}
\and E. Carolo\inst{2,6} 
\and S. Chinellato\inst{2,6} 
\and J. Farinato\inst{2,6} 
\and D. Magrin\inst{2,6} 
\and L. Marafatto\inst{2,6}
\and G. Umbriaco\inst{1,4,5,6}
\and D. Vassallo\inst{2,6}
}

\institute{Dipartimento di Fisica e Astronomia ``G. Galilei'', Universit\`a di Padova, vicolo dell'Osservatorio 3, I-35122 Padova, Italy\\
\email{roberto.ragazzoni@unipd.it}
\and INAF-Osservatorio Astronomico di Padova, vicolo dell'Osservatorio 5, I-35122 Padova, Italy
\and INAF-Osservatorio Astronomico d'Abruzzo, via Mentore Maggini snc, I-64100 Teramo, Italy
\and Dipartimento di Fisica e Astronomia "A. Righi" - Alma Mater Studiorum Universit\`a di Bologna, via Piero Gobetti 93/2 - 40129, Bologna, Italy
\and INAF-Osservatorio di Astrofisica e Scienza dello Spazio, via Gobetti 93/3, 40129, Bologna
\and ADONI - Laboratorio Nazionale Ottiche Adattive - Italy
}

\date{Received ...; accepted ...}
\abstract
% 5 {} token are mandatory
  % context heading (optional) {.} %leave it empty if necessary
  {Full sky coverage Adaptive Optics on Extremely Large Telescopes requires the adoption of several Laser Guide Stars as references. With such large apertures, the apparent elongation of the beacons is absolutely significant. With few exceptions, WaveFront Sensors designed for Natural Guide Stars are adapted and used in suboptimal mode in this context.}
  {We analyse and describe the geometrical properties of a class of WaveFront Sensors that are specifically designed to deal with Laser Guide Stars propagated from a location in the immediate vicinity of the telescope aperture.}
  {We describe in three dimensions the loci where the light of the Laser Guide Stars would focus in the focal volume located behind the focal plane (where astronomical objects are reimaged). We also describe the properties of several types of optomechanical devices that, through refraction and reflections, act as perturbers for this new class of pupil plane  sensors, which we call ingot WaveFront Sensor.}
  {We give the recipes both for the most reasonable complex version of these WaveFront Sensors, with 6 pupils, and for the simplest one, with only 3 pupils. Both of them are referred to the ELT case.
   Elements to have a qualitative idea of how the sensitivity of such a new class of sensors compared to conventional ones are outlined.
  }
  {We present a new class of WaveFront Sensors, by carrying out the extension to the case of elongated sources at finite distance of the pyramid WaveFront Sensor and pointing out which advantages of the pyramid are retained and how it can be adopted to optimize the sensing.}

 
\titlerunning{An Ingot-like class of WFSs for LGSs}

  \authorrunning{R. Ragazzoni et al. }

   \maketitle
%
%________________________________________________________________

\section{Introduction}
\label{sec:intro}
Adaptive Optics (AO) allows achieving diffraction
limited imaging on large ground-based telescopes \citep{Tyson1991,Beckers1993,Hardy1998} and can be characterized, in terms of achieved performances, by three parameters. The first one represents the degree of the achieved compensation, often indicated by the ratio of the peak of the compensated Point Spread Function of an unresolved source to the diffraction limit one, the so-called Strehl ratio \citep{Strehl1895,Herrmann1992}. This compensation ranges from seeing amelioration \citep{Rigaut2002,Tokovinin2004}, to modest Strehl ratios, up to the extreme AO case that is encompassed by high-contrast system usually aiming at the detection of circumstellar or exoplanets around bright stars \citep{Macintosh2001, Esposito2003, Esposito2010, Guyon2018} and limited to a very narrow Field of View (FoV). The second parameter is, in fact, the size of such a compensated FoV that is naturally limited to the isoplanatic patch in the case where a single conjugated AO is achieved. Multiple correctors in a large variety of configurations both for compensation and sensing have been described  \citep{Dicke1975, Beckers1988, Ellerbroek1994, Ragazzoni2000,Ragazzoni2002, Ragazzoni2014, Rigaut2020} and some tested on sky to date \citep{marchetti_mad_2003, Marchetti2008,Rigaut2014, Neichel2014, Herbst2018}. Finally, the third fundamental parameter is the sky-coverage, the fraction of the sky where the compensation can actually be achieved.

The sky coverage issue has been addressed by conceiving AO systems that can work with rather faint reference sources, eventually scattered in a large area in the sky \citep{Ragazzoni2013, Viotto2015, Portaluri2017}, or by producing artificial reference sources by propagating light beams from the ground \citep{Thompson1987,Rigaut2014} and sensing the signal provided by some of the returning light by means of different processes, such as Rayleigh \citep{Foy1985} or resonant scattering \citep{Pilkington1987}. 

Sodium Laser Guide Stars (LGSs) became routinely available in a few observatories \citep{Rigaut2014, Calia2014, DOrgeville2016} also in multiple formats (possibly forming true artificial asterisms) and are usually considered as the obvious solution to the sky coverage issue. While the AO system that first reliably and consistently used these artificially generated beacons succeeded in producing unique science \citep{Genzel2003,Ghez2008}, it is remarkable to note that anyhow they would rely on some Natural Guide Stars (NGSs) because of indetermination of the absolute position in the sky of the artificial reference due to the upward turbulence encountered by the light  propagating from the ground to the sky \citep{Rigaut1992, Foy1995,Ragazzoni1996, Ragazzoni1997, Esposito2000, Ragazzoni2000nato}.

Assuming the avalaibility of the desired return flux from these LGSs, their associated WaveFront Sensor (WFS) is not necessarily required to be characterized by any minimum efficiency. In fact, the current examples of AO systems encompassing LGSs are just conventional kind of WFSs, usually conceived for NGSs, adapted to the location and wavelength of these peculiar references.

However these sources -from an optical viewpoint- in spite of being nicknamed as "stars" are all but unresolved sources located at an infinite distance as an NGS is. Further to the obvious evidence that they are located at a finite distance (typically at 92~km of height, corresponding to a continuously changing range that depends upon the zenith distance at which the observation is carried out), they have a non-negligible distribution in all the spatial dimensions, and especially along the line of propagation of the laser, not necessarily co-aligned with the line of sight.

The departure of the appearance of the LGSs from an NGS (provided proper refocussing is done) is somehow proportional to the diameter of the observing telescope so that this issue has minimal impact on current 8 to 10~m state of the art class telescopes, but it is expected to play a more significant role in the future gigantic ground based optical telescopes like the European Extremely Large Telescope (ELT, \citealt{Gilmozzi2007}), the Giant Magellan Telescope (GMT, \citealt{Johns2008}), and the Thirty Meter Telescope (TMT, \citealt{Szeto2008}). Two of the most prominent instruments on those telescopes will make use of LGS adopted Shack-Hartmann (SH) WFS, as a well consolidated, although probably suboptimal, technology \citep{Boyer2016,Agapito2022}.

The paper is organized as follows.
Sect.~\ref{sec:problem} describes the geometry of the LGS, including the relationships  between the source and the elongated spot on the focal volume. Sect.~\ref{sec:newclass} explores the several possible options for the new class of WFSs, combining reflecting and refracting surfaces. It provides a geometric description of the WFS (Subsect. \ref{subsec:geom}) and focuses on two extreme realizations with 6 (Subsect.~\ref{subsec:6pup}) and 3 pupils  (Subsect.~\ref{subsec:3pup}).
Sect.~\ref{sec:results} resumes and concludes the work.

%__________________________________________________________________

\section{Description of the LGS in the focal volume}
\label{sec:problem}
Treating the LGSs as a cylindrical emitting volume rather than point sources reveals several suboptimal characteristics in the associated WFS, leading to an increase in the required returning flux to attain a certain level of accuracy.

Most of the literature dealt with the apparent angular spreading of LGSs (called elongation), managing a number of countermeasures or specific technical issues, such as the truncation of its image when treated by a classical WFS \citep{Diolaiti2012,Schreiber2014,Vieira2018,Clare2020}.

However, in principle, the conceptual design of a WFS could incorporate the way this light is produced, tailoring that specific need, and thus being more efficient than treating LGSs as stars \citep{Ragazzoni2017}. Some of these approaches have been described in the literature: the z-invariant WFS \citep{Ragazzoni2001}, the PPPP concept \citep{Buscher2002,Yang2018}, SPLASH \citep{SPLASH}, the CAFADIS \citep{Ramos2008}, the Plenoptic WFS \citep{Zhang2021} or through a smart optical arrangement of the SH beams \citep{Lombini2022}. Some configurations have been arranged on a laboratory breadboard where the overall configuration is scaled down to a laboratory size \citep{Yang2019}. Finally, some of them have been even tested, although with some limitations \citep{Bharmal2018}, on the sky (\citealt{Ragazzoni2006}).

We extend now one of these earliest concept, namely the roof-like WFS \citep{Ragazzoni2001b}, customizing it to the Sodium LGSs, and extending its ability to provide information in both the directions: the orthogonal one and the one aligned with respect to the apparent elongation of the Sodium beacon as seen from the main aperture of the telescope.

\begin{figure}
    \centering
    \includegraphics[width=9cm]{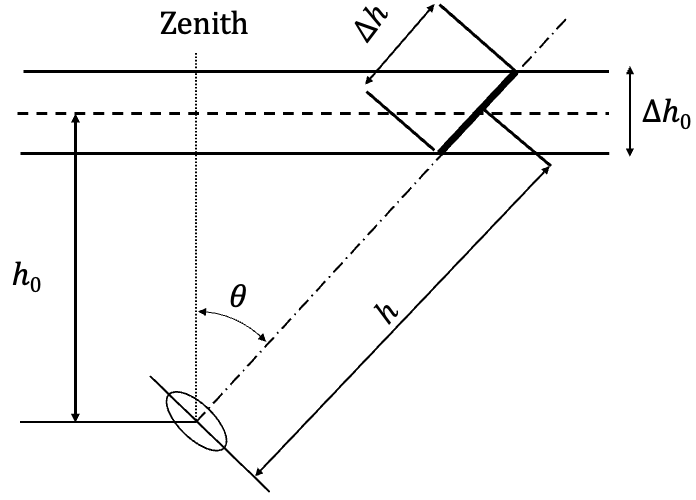}
    \caption{The geometry of the launching of a mesospheric Sodium layer LGS. In this schematic view, the layer is approximated with a solid bandwagon of thickness $\Delta h_0$ located at an averaged height of $h_0$. Curvature of the Earth is considered negligible, making the zenith distance $\theta$ the only parameter to retrieve the equivalent range $h$ and the apparent thickness $\Delta h$.}
    \label{fig:geometryNa}
\end{figure}

Let us consider a laser tuned to the proper Sodium doublet wavelengths propagating the beam through a projector producing an LGS at a zenith distance $\theta$ (Fig.~\ref{fig:geometryNa}). The Sodium layer, located at an altitude of $h_0$ and nominal thickness $\Delta h_0$ will be seen from the launching area at a range $h$ and with an equivalent projected thickness $\Delta h$ given by:
\begin{equation}
\centering
    h=\frac{h_0}{\cos{\theta}}; \hspace{1cm} \Delta h =\frac{\Delta h_0}{\cos{\theta}}.
\end{equation}

Each point of the Sodium column emitting light will appear with an apparent defocus if observed on the nominal focal plane at which objects at infinity are conjugated.

If the equivalent focal length of the telescope is $f$, $F$ is the focal ratio of the optical system and $D$ is its aperture,  one can note that a point $C$ located in the nominal range at the Sodium layer, will be reimaged well behind the focal plane where astronomical objects are focused (Fig.~\ref{fig:geomdim}). The detailed position is indicated by $h'$ and $s'$ under the conditions that the LGS is being propagated from a projector located at a distance $s$ from the center and on the plane of the telescope pupil $O$, and -only for the sake of this preliminary discussion- the propagation of the LGS is parallel to the optical axis of the telescope. In this configuration, and using the thin lenses approximation, one can figure out the distance along the optical axis of the reimaged position of the point $C'$. The axial distance is given by:

\begin{figure}
    \centering
    \includegraphics[width=12cm,angle=90]{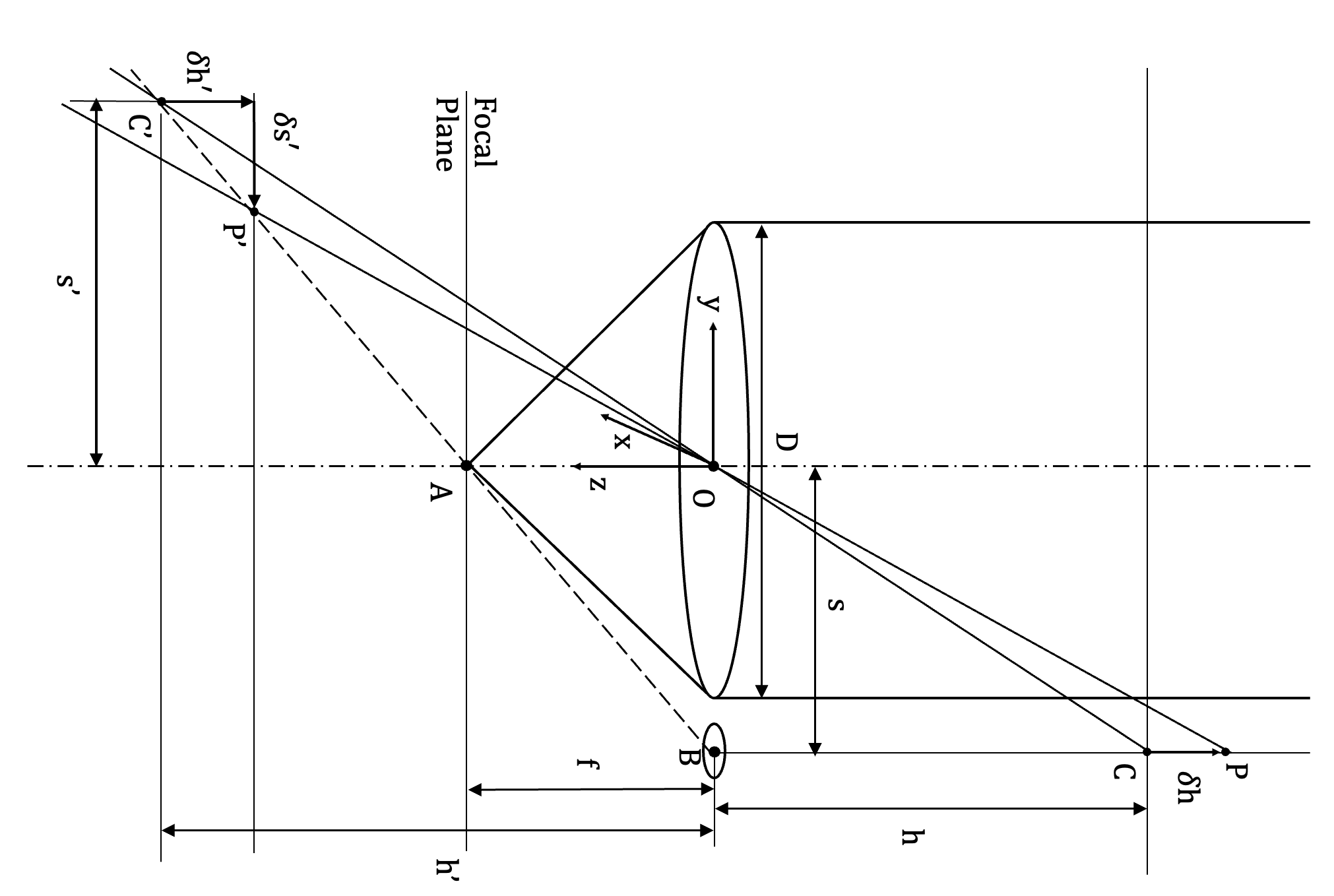}
    \caption{Geometry describing the reimaging of a vertically extended LGS beacon projected from outside the telescope pupil and co-aligned with its optical axis. Points $P$ and $C$ of the LGS are reimaged onto $P'$ and $C'$ that are aligned with $A$ and $B$, respectively the focus of an on-axis astronomical source and the center of the LGS projector.}
    \label{fig:geomdim}
\end{figure}

\begin{equation}
h'=\frac{fh}{{h-f}}.
\label{eq:thinlens}
\end{equation}

This relationship, whenever $f \ll h$ (a condition generally well accomplished), can be approximated by the following:

\begin{equation}
h' \approx f + \frac{f^2}{h}.
\label{eq:thinlensapprox}
\end{equation}

Similarly the lateral displacement with respect to the optical axis of the imaged point $C'$ can be figured out by using the similarity of the proper triangles having a common vertex in the center of the telescope pupil $O$, through the following relationship:

\begin{equation}
\frac{s'}{h'} = \frac{s}{h},
\label{eq:triang}
\end{equation}

that fully characterizes the position of $C'$.

As the LGS extends along the direction of propagation of the laser beam, we now work out the effects of a relatively small displacement from the starting point $C$, namely to a generic point $P$ located at a displaced distance from the laser beam by a (small) amount $\delta h$ typically lying in the range $\pm\Delta h/2$.

In particular, one can rewrite eq.~\ref{eq:thinlens} by replacing the displaced position of the reimaged point $P'$  as:
\begin{equation}
h'-\delta h' = \frac{f(h+\delta h)}{h+\delta h - f}.
\label{eq:puntoP}
\end{equation}

Under the reasonable assumption that both $f,\delta h \ll h$ and using also eq.~\ref{eq:thinlensapprox}, one can rewrite it into the following:

\begin{equation}
\frac{\delta h}{\delta h'} \approx \frac{h^2}{f^2}.
\label{eq:puntoPapprox}
\end{equation}

One can obtain an equivalent relationship to eq.~\ref{eq:triang} using the similarity of the angles having as vertexes $P$ and $P'$ and, again, common vertex in the center of the telescope pupil, as:

\begin{equation}
\frac{s'-\delta s'}{h'-\delta h'} = \frac{s}{h+\delta h}.
\label{eq:triangP}
\end{equation}

After a tedious but straightforward rearrangement using sequentially the results given by  eqs.~\ref{eq:puntoP},~\ref{eq:puntoPapprox}, \ref{eq:thinlens}, and adopting the mentioned approximation, one can finally figure out the following:

\begin{equation}
\frac{\delta s'}{\delta h'} \approx \frac{s}{f}.
\label{eq:finalgeomdem}
\end{equation}

It is interesting that this result indicates that the locus of the points where the image of the elongated LGS is lying can be described as a segment. Such pattern is inclined with respect to the optical axis in a manner that, if prolonged, will hit both the nominal focus of an infinite distant on axis source and the position, along the plane of the pupil, where the beam is propagated (see the dashed line in Fig.~\ref{fig:geomdim}) . In different words, within such an approximation, this finding leads to the consequence that the triangles with hypotenuses $\overline{C'P'}$ and $\overline{AB}$ are similar. This is the so-called Scheimpflug principle \citep{ScheimpflugBrevetto, Scheim1994}, which can be geometrically explained by extending the power of the telescope outside of its nominal diameter up to reach the LGS projector and using the thin lenses geometric approximation to figure out the reimaged points along the LGS beam. 

\begin{figure}
    \centering
    \includegraphics[width=8.7cm]{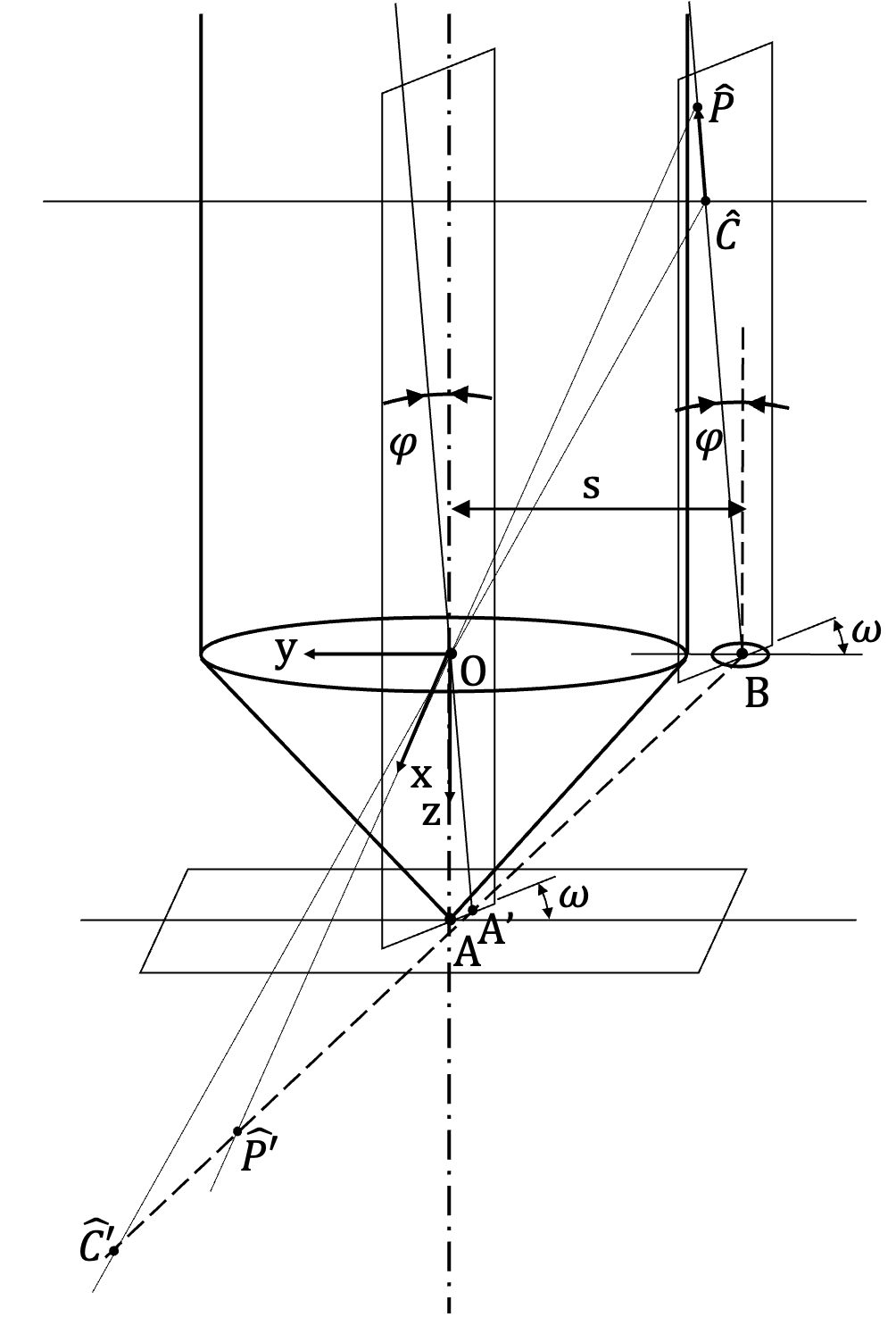}
    \caption{The Scheimpflug principle applied to a generic position, defined by the angles $\varphi$ and $\omega$, and line of sight of the LGS projector. In this case, the alignment between $\hat{C'}$ and $\hat{P'}$ and $B$ intersects the nominal focal plane for astronomical objects in the point $A'$. This is where an astronomical source located in the direction of the LGS will focus.} 
    \label{fig:Splane}
\end{figure}

The Scheimpflug principle can be applied to any direction of propagation of the LGS angles $\varphi$ and $\omega$, as shown in Fig.~\ref{fig:Splane}. 

As a consequence, when the LGS projector is located outside of the telescope pupil, the illuminating beams will never embed the LGS image itself.
On the contrary, in the case of a LGS propagated from behind the cage of a secondary mirror, in a typical two-mirrors telescope, the locus where the LGS image is  in focus lies within the beams coming from other portions of the LGS. This has nothing to do with the fratricide effect \citep{Gratadour2010}, which is related to the augmented background due to the light scattered from the LGS itself.
For this reason, the class of WFSs presented in this paper is suitable only to those telescopes in which the LGS projector is placed outside of the telescope pupil as will be evident in the following section.

\section{Toward a new class of WFSs}
\label{sec:newclass}

The whole light emitted by the LGS can be subdivided in different regions, with the aim of gathering information on how the light has been perturbed by the atmosphere. 
This can be accomplished by reflecting or refracting interfaces that would bend the incoming beam, to be then recollected by a common pupil imager, to form a sort of pupil-plane WFS tailored to the specific case of LGSs. 
 This type of WFSs gained an increased attention in the recent decades, mainly because of the demonstrated high sensitivity. They include an optomechanical device located in the focal volume and a reimager of the pupil plane. The term "pupil-plane" refers to the location where the detector is placed (see for instance \citealt{Horwitz1994} and \citealt{Ma2016}). A detailed description and a list of this category of WFSs can also be found in \cite{Ragazzoni2019}.

Because of the solid shape, in this case such a reflective and refractive device combination resembles the surfaces of an ingot, and thus the nickname we give to this class of WFSs.
In the following, we establish a coordinate system which origin $O$ is located in the center of the entrance pupil of the main telescope; the $z$ axis deploys along its optical axis being positive toward the focal plane; the $y$ axis defines the location where the LGS projector is located (at position $y=-s$) such that, generally speaking, the image of the LGS falls at positive values of $y$; finally the $x$ axis is added in order to be arranged in a normal Cartesian coordinate system.  

In a similar fashion to how different pupils illumination are used to sense the derivative of the wavefront \citep{Ragazzoni1996b} in two dimensions in the pyramid WFS, these multi-faceted prisms would produce a corresponding number of pupil images to be used for similar purposes.
In this way, one can efficiently measure the derivative of the wavefront along the $x$-axis by combining all the pupils collecting the light from the whole of the LGS, and the derivative of the wavefront along the $y$-axis using the light coming from the endpoints of the LGS.
A crude version of this approach, suited to sense solely the derivative along the $x$-axis of a Rayleigh reference beacon, appeared in \cite{Ragazzoni2000b}. In that context, no attempt was made for measuring the derivative along the $y$-axis, given the fact that such references are characterized by a vanishing or non existing edge along the direction of propagation.

The case of the conventional pyramid WFS used with LGS is not treated in details here, however, as it deals with sources much larger than the diffraction limit ones in all dimensions (including the one orthogonal to the elongation), it can be regarded as a sub case of a SH with a 2x2 pixels per subaperture, hence with a very large oversampling of the LGS image.
It is also fair to point out that one of the key features of the pyramid WFS, namely the enhanced sensitivity reached in closed loop \citep{Ragazzoni1999}, is not achievable here.
This is because the LGS is usually much larger than the diffraction limit of the telescope so that closed loop operations do not change significantly its apparent size and cannot be used to retrieve any advantage.
The only way to retain such a feature would be by producing an LGS whose apparent size, other than the elongation, is comparable to the diffraction limit of the telescope. This can be accomplished by propagating the LGS through a similarly sized aperture, with the additional incumbency of compensating for the upward atmospheric turbulence  \citep{Esposito2008}. Practically this can be obtained efficiently only using the same telescope aperture as a laser projector, with the obvious associated complexity of achieving such a task \citep{Fugate1994, Wilson1997}. 
For these reasons, such a possibility is not further elaborated in this work.

To obtain a similar behaviour on a SH WFS, one should take provisions for the largest number of pixels required at the subapertures more distant from where the LGS is being propagated. This would require an overall very large number of pixels that would be almost unused in the regions close to where the LGS is being propagated, where the apparent elongation is small.

The overall number of regions subdividing the LGS image each (producing a pupil image) can be, in principle, as large as desired, provided that practical issues are taken into account. For instance, a minimum requirement for this approach is that the bending angles are such that the commonly reimaged pupils do not superimpose. This fact translates into the condition that the bending angles are at least the angle defined by the marginal rays of the converging beams, amounting to the inverse of the focal ratio at the LGS image position (Fig.~\ref{fig:chiefray}).
Moreover, a large number of pupil images will require larger bending angles, corresponding to a larger FoV capability of the collimating system.

We believe that, once one gives up using the structure of the brightness along the LGS, the maximum reasonable number of regions is six \citep{Ragazzoni2018}. This is the case depicted in Fig.~\ref{fig:chiefray} and in the first case of Table~\ref{tab:variingot}. 

There are, however, a number of interesting features that are common to all the versions of WFSs that are described in this work. The distribution of light in points of the pupil that experience different elongations reflects the availability of the light for sensing the derivative of the wavefront in different directions. In other words, points closer to the location where the LGS is propagated will see the WFS as a nearly pyramid one (with just, or mainly, four faces in the most complete case). Instead, when the largest elongation occurs, the largest fraction of the light is only used for sensing in the $x$-axis and just the residual fraction of the light on the edges is used for sensing along the $y$-axis. 

It should be noted that any significant structure in the Sodium column density (for example caused by a sudden bright layer, as occasionally experienced) will be smeared out, by a significant angular width of the upwarding beam.
The six regions approach, in fact, deliberately does not make use of the eventual structures seen in the elongated LGS, as per different densities of the Sodium layer. 

Asymmetric solutions along the $y$-axis should be taken into consideration as well, recalling that the physical distribution of the Sodium layer is generally asymmetric. Assuming that the layer "floats" above a certain height one can expect that the lower edge of the beacon will be sharper than the upper one \citep{Ragazzoni2018}. This would point toward solutions with a limited number of slicing regions, down to a minimum of three, \citep{Ragazzoni2019}. In this configuration, only what is expected to be the sharpest edge is used to sense the derivative of the wavefront along the $y$-axis. 

Fig.~\ref{fig:tel+ingot} shows a possible arrangement for a 6-pupils ingot, but a lot of other options are possible, as results from different combinations of surfaces: only reflective, or only refractive, a combination of both, depending on which region of the LGS image is being considered.
The possible combinations of reflective and refractive surfaces have mainly implication of practicality. Provided that the inclination of the LGS focus on the focal volume is typical a small perturbation with respect to the optical axis, making this surface refractive an unpractical solution, the adoption -for instance- of only reflective surfaces would lead to the use of the collimator mainly off-axis. In the configurations we adopted, we qualitatively choose combinations that lead to chief rays of the various subapertures elongating into opposite directions with respect to the optical axis. This choice is to make less demanding the specifications on the quality of the collimating optics in terms of useful FoV.

However, for any reasonable choice of splitting regions, the overall demand in terms of pixel format on the CCD is vastly smaller than any SH WFS \citep{Ragazzoni2019} with comparable spatial resolution on the WF.
Considering a squared detector, let us note as $N_p$ its side size in pixels. For a 6-pupils configuration, arranged in a $2\times 3$ format, one can easily estimate:
\begin{equation}
    N_p \sim 3N ,
\end{equation}
where $N$ is the number of subapertures along the diameter.
The same estimation can be done for a SH WFS, by quantifying the largest elongation $\Delta\epsilon$ observed from the edge of the pupil opposite to the LGS launcher: 
\begin{equation}
\Delta\epsilon \approx \frac{\Delta h}{h} \cdot \frac{D/2+ s}{h} \le \frac{\Delta h_0}{h_0^2}\cdot D,
\end{equation}

where the upper limit, happening for zenith observations ($\theta=0$)
is calculated for the most favourable condition for the position of the projector ($s=D/2$).

If we consider the minimum apparent diameter of a non elongated LGS propagated through a projector with perfect optical quality as $\sqrt{2}\epsilon_s$ (with $\epsilon_s$ being the seeing angle) and assuming a Nyquist sampling, a minimum subaperture size in pixels is given by: 
\begin{equation}
\Delta_p =\sqrt{2}\frac{\Delta\epsilon}{\epsilon_s}.
\end{equation}
Therefore, the overall detector format must be at least \begin{equation}
N_p \sim N\Delta_p.
\end{equation}

Assuming $D=40$ m, $\Delta h_0=20$ km and $N=100$ one gets, for the ingot case an $N_p \approx 300$ with respect to $N_p\approx 2100$ for the SH case, leading to a size ratio of $\approx 7$ times. This should be retained as a conservative estimate in favor of the ingot concept.
In fact, such a ratio scales linearly with $\Delta h_0$ and increases as much as the LGS is being propagated away from the edge of the primary mirror, as any practical case will be.

It is interesting to point out that smaller detector formats are possible, as depicted in \cite{Agapito2022} where with $D=39$ m, $N=70$ and $N_p=1100$ are being adopted. This, however, is done at the expenses of handling in some manner an incomplete sampling of the LGS (truncation).

As a last consideration, one should take into account that, because of the cone effect, it is mandatory to use this WFS scheme in a MCAO or similar system, such that the information per subaperture could come from different beacons. This means that a simple roof solution (although not further explored here) could give satisfactory results with a large enough number of LGSs, in spite of being unsatisfactory to solve completely the wavefront information from a specific beacon. 

\subsection{Geometrical description}
\label{subsec:geom}

\begin{figure}
    \centering
    \includegraphics[width=5 cm]{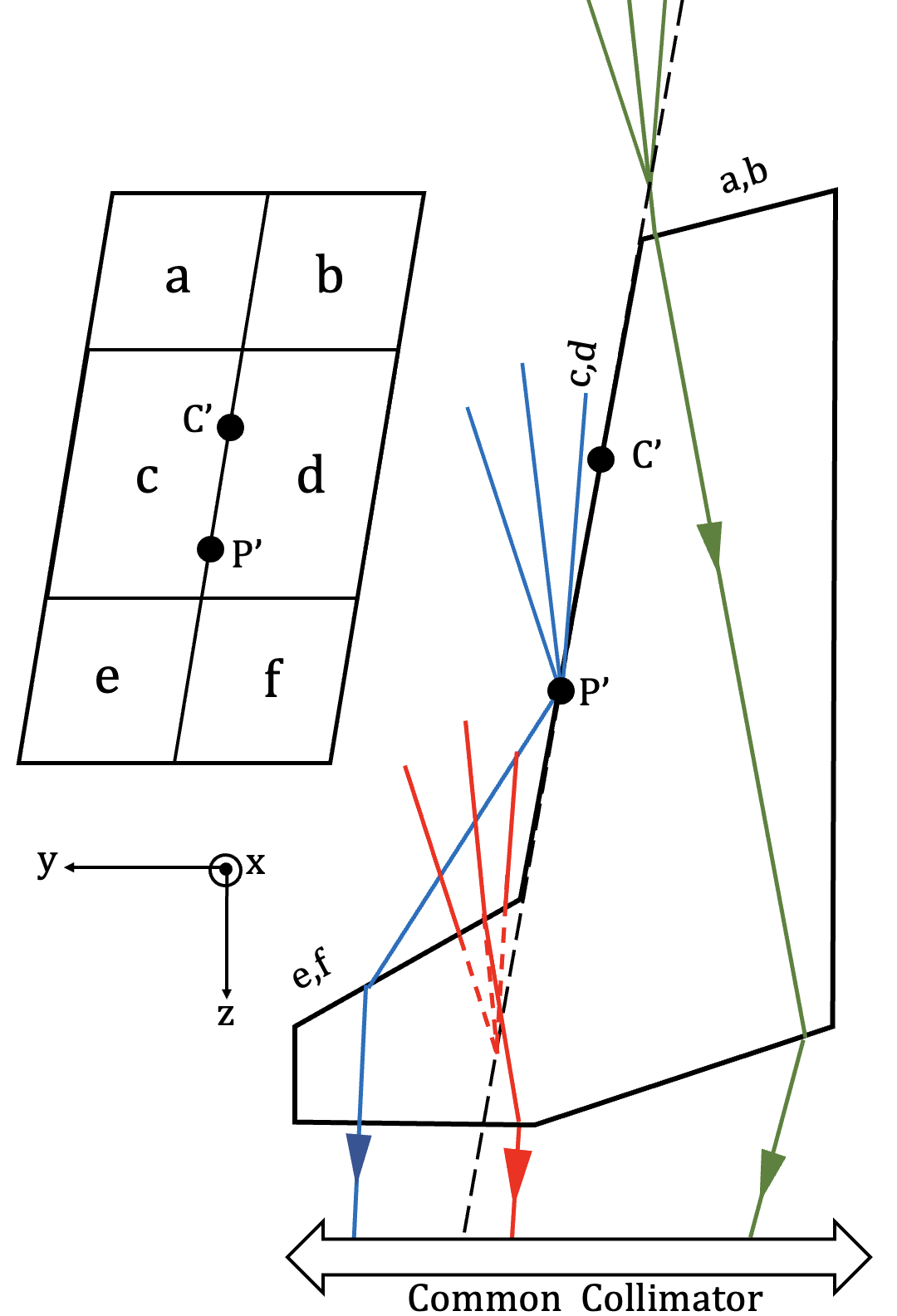}
    \caption{Chief rays propagating through the 6 faces. A view from above reproducing the lettering of the surfaces is given in the upper left side.}
    \label{fig:chiefray}
\end{figure}

\begin{figure}
    \centering
    \includegraphics[width=9cm]{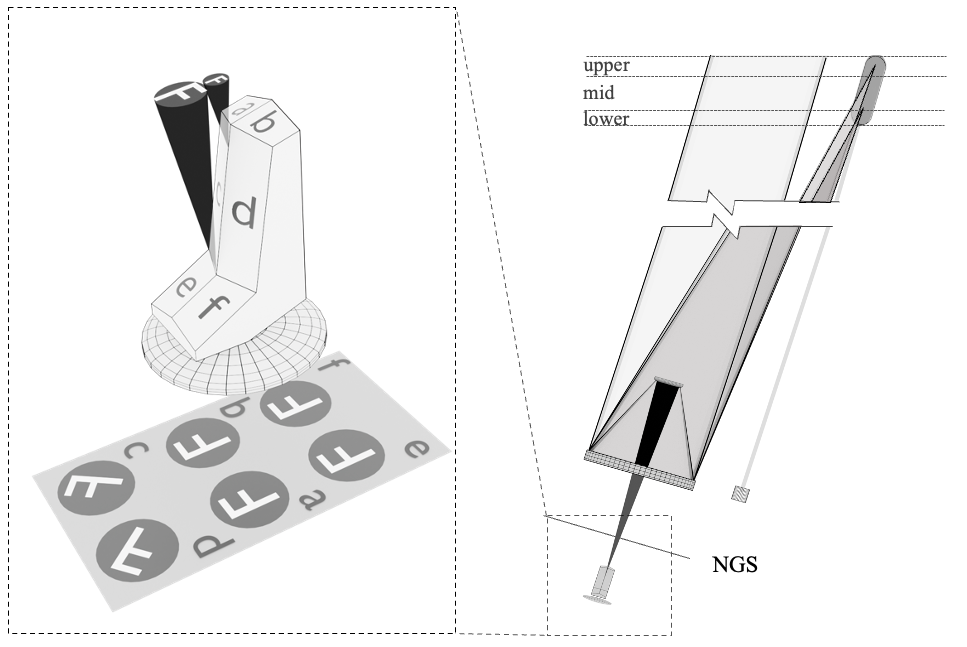}
    \caption{In this illustrative picture the telescope on the right is shown together with the conjugation of the various layers of the Sodium LGS beacon. On the zoomed portion portrayed on the left side, the beams coming from the upper and lower portions of the artificial reference are pointed out together with the pupil rotations highlighted by the "F" on the footprint of the pupils, because of the reflections and refractions through the ingot.}
    \label{fig:tel+ingot}
\end{figure}

\begin{figure*}
    \centering
    \includegraphics[width=18cm,height=9.5cm]{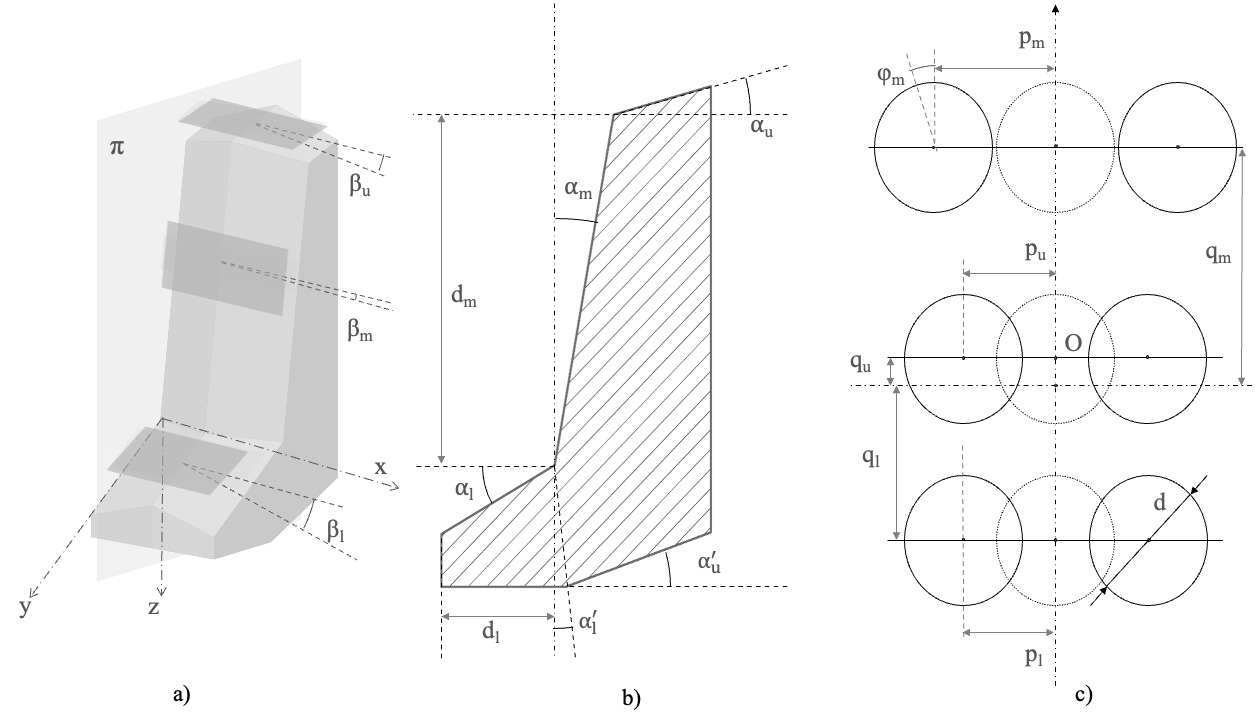}
    \caption{( a): Ingot prism solid shape with 6 faces. (b): Section of the prism. (c): 6 pupils corresponding to the 6 faces. $u,m,l$ refer to upper, medium and lower portion of the ingot.} 
    \label{fig:3piani}
\end{figure*}

Let us consider the ray-tracing equations to describe the ray deviations operated by the faces of the ingot. When a ray with direction $\hat{r}=(r_x;r_y;r_z)$ is reflected by a mirror with the normal to the surface having direction $\hat{n}=(\hat{n}_x;\hat{n}_y;\hat{n}_z)$, the direction of the reflected ray $\hat{r'}$ is (\citealt{2012TextbookLaser}):  
\begin{equation}
\hat{r'} = M \cdot \hat{r},
\end{equation}
where the reflection matrix M is:
\begin{equation}
M = I - 2\hat{n} \cdot \hat{n}^T ,
\end{equation}
where $I$ is the canonical identity matrix.
The law of refraction can also be expressed in vector form and the direction of the refracted ray $\hat{r'}$ is (\citealt{2012TextbookLaser}):
\begin{equation} \label{eq:refraction}
\hat{r'}=\left\{\sqrt{1-\mu^2\left[1-\left(\hat{n}\cdot \hat{r}\right)^2\right]}-\mu\left(\hat{n} \cdot \hat{r}\right)\right\}\hat{n}+\mu \hat{r},
\end{equation}
where:
\begin{equation}
\mu = \frac{n_1}{n_2}
\end{equation}
is the ratio between the refraction indices for the incident (incoming) and transmitted (outcoming) mediums.

Let us now consider the case of a telecentric telescope with a chief ray propagating along the z direction: $\hat{r} = (0; 0; 1)$.
We consider a mirror oriented perpendicularly to the incoming ray: $\hat{n} = (0; 0;-1)$.
We rotate the mirror around $x$ by an angle  $\theta_x$ and we further rotate it around $y$ by an angle $\theta_y$. By applying the double rotation, the final orientation of the mirror will be: 
\begin{equation}
\hat{n}'  = R_x \cdot R_y \cdot \hat{n} = \left( \begin{matrix} 
\sin{\theta_y} \\ -\cos{\theta_y}\sin{\theta_x} \\- \cos{\theta_y}\cos{\theta_x}\end{matrix} \right), \\
\end{equation}
where $R_x$ and $R_y$ are the rotation matrices around the x and y axes respectively. The reflection matrix associated with it is:
\begin{equation}
\begin{tiny}
M= \left( \begin{matrix} 
1-2 \sin^2{\theta_y}                           & 2 \sin{\theta_y} \cos{\theta_y} \sin{\theta_x} & 2 \sin{\theta_y} \cos{\theta_y} \cos{\theta_x} \\
2 \sin{\theta_y} \cos{\theta_y} \sin{\theta_x} & 1-2 \sin^2{\theta_x}\cos^2{\theta_y}  & - 2\cos^2{\theta_y} \sin{\theta_x} \cos{\theta_x} \\
2 \sin{\theta_y} \cos{\theta_y} \cos{\theta_x} &  - 2\cos^2{\theta_y} \sin{\theta_x}\cos{\theta_x} &  1-2 \cos^2{\theta_y}\cos^2{\theta_x} 
\end{matrix} \right).
\end{tiny}
\end{equation}
The orientation of the ray after the reflection off the mirror is:
\begin{equation}
\hat{r'}= \left( \begin{matrix} 
2 \sin{\theta_y} \cos{\theta_y} \cos{\theta_x} \\
-2 \cos^2{\theta_y} \sin{\theta_x} \cos{\theta_x}\\
1-2 \cos^2{\theta_y} \cos^2{\theta_x}  
\end{matrix} \right) =
\left( \begin{matrix} 
\sin{2\theta_y} \cos{\theta_x} \\
-\sin{2\theta_x} \cos^2{\theta_y}\\
1-2\cos^2{\theta_y} \cos^2{\theta_x}
\end{matrix} \right).
\end{equation}
Referring to the upper, medium and lower part of the ingot as $u,m,l$, and calling $\alpha_m=90^{\circ}-\theta_x$ and $\beta_m = \theta_y$, as shown in  part (a) and (b) of Fig.~\ref{fig:3piani}, we can rewrite the orientation of the reflected ray as:
\begin{equation}\label{eq:reflected_ray}
\hat{r'}= 
\left( \begin{matrix} 
\sin{2\beta_m} \sin{\alpha_m} \\
-\sin{2\alpha_m} \cos^2{\beta_m}\\
1-2\cos^2{\beta_m} \sin^2{\alpha_m}
\end{matrix} \right).
\end{equation}
In the approximation of small angles, equation~\ref{eq:reflected_ray} becomes:
\begin{equation}\label{eq:small_angle_refl}
\hat{r'}\approx 
\left( \begin{matrix} 
2\beta_m \alpha_m \\
-2\alpha_m\\
1-2\alpha_m^2
\end{matrix} \right),
\end{equation}
where such small angles are expressed in radians.
It is worth to note that the central pupils (regions c and d in Fig.~\ref{fig:chiefray} and \ref{fig:tel+ingot}), due to the reflection, are flipped and rotated of an angle $\varphi_m$ given by:
\begin{equation}\label{eq:flippupils}
\varphi_m \approx 2\beta_m \cos\alpha_m.
\end{equation}
This is just a feature since they can be calibrated with an interaction matrix considering, for instance, modal bases.

A similar approach can be used for the refractive faces. In this case, taking as an example the lower part of the ingot, characterized by the angle $\alpha_l$, as indicated in Fig.~\ref{fig:3piani}, we can proceed as follows: 
\begin{itemize}
\item start from a refractive plane perpendicular to the $z$-axis: $\hat{n} = (0;0;1)$
\item apply a rotation around $x$ of $\theta_x = \alpha_l$ and a rotation around $y$ of $\theta_y = \beta_l$ to find the orientation of the normal vector. 
\item apply equation~\ref{eq:refraction} consecutively for each refracting plane to calculate the ray vector exiting the ingot. 
\item in the case of small angles the deviation of the ray can be expressed with the following simple relation: 
\begin{equation}\label{eq:small_angle_refr}
\hat{r'}\approx 
\left( \begin{matrix} 
\left(n-1\right)\beta_l\\
\left(n-1\right)\alpha_l\\
1-0.5\left(n-1\right)\left(\beta_l+\alpha_l\right)
\end{matrix} \right).
\end{equation}
\end{itemize}

Note that, at first approximation, $\beta_l$ and $\alpha_l$ are respectively the $x$ and $y$ component  of the apex angle of the prism constituting the bottom part of the ingot. The first two components of $\hat{r'}$ are basically the deviations operated by a thin prism with refractive index n, while the third component is the approximation of the normalization factor.

Using equations \ref{eq:small_angle_refl} and \ref{eq:small_angle_refr}, we can express the position of the pupils with respect to the optical axis (see Fig.~\ref{fig:3piani} part c) when they are reimaged by a lens with focal length $f_c$:

\begin{equation}
\begin{gathered}\label{eq:6pup}
\begin{split}
p_m &\approx f_c [2\beta_m \alpha_m + (n-1)\beta_l] ;& q_m & \approx f_c [2\alpha_m - (n-1)\alpha_l];\\
p_u &\approx f_c  (n-1)\beta_u ;& q_u &\approx f_c (n-1)(\alpha_u -\alpha'_u);\\
p_l &\approx f_c (n-1)\beta_l ;& q_l &\approx f_c (n-1)\alpha_l.\\
\end{split}
\end{gathered}
\end{equation}

Recalling that $d\approx f_c/F$, one can use, as a first approximation, the above equations to properly fit pupils into the chosen detector format in order to avoid overlaps, and keeping a minimum guard distance among pupils to prevent scattering light from one pupil to the other. Those equations can be also used, of course, to proper engineer the detailed shape of the ingot in order to minimize deviations of the chief rays and second order effects like distortions of the pupil. 
It is worth to point out that a detailed raytracing is needed in order to refine the final layout and given the monochromatic nature of the LGS, chromatism is not an issue at any degree of approximation, in contrast with the case of the pyramid WFS used with NGSs \citep{Diolaiti2003}.
Negative values of the angles considered here should be taken into account in this optimization process that depends upon the particular focal ratio of the chosen telescope. Furthermore, the non telecentricity of the telescope is to be properly included, at least in the final ray tracing analysis.

\begin{figure}
    \centering
    \includegraphics[width=9cm]{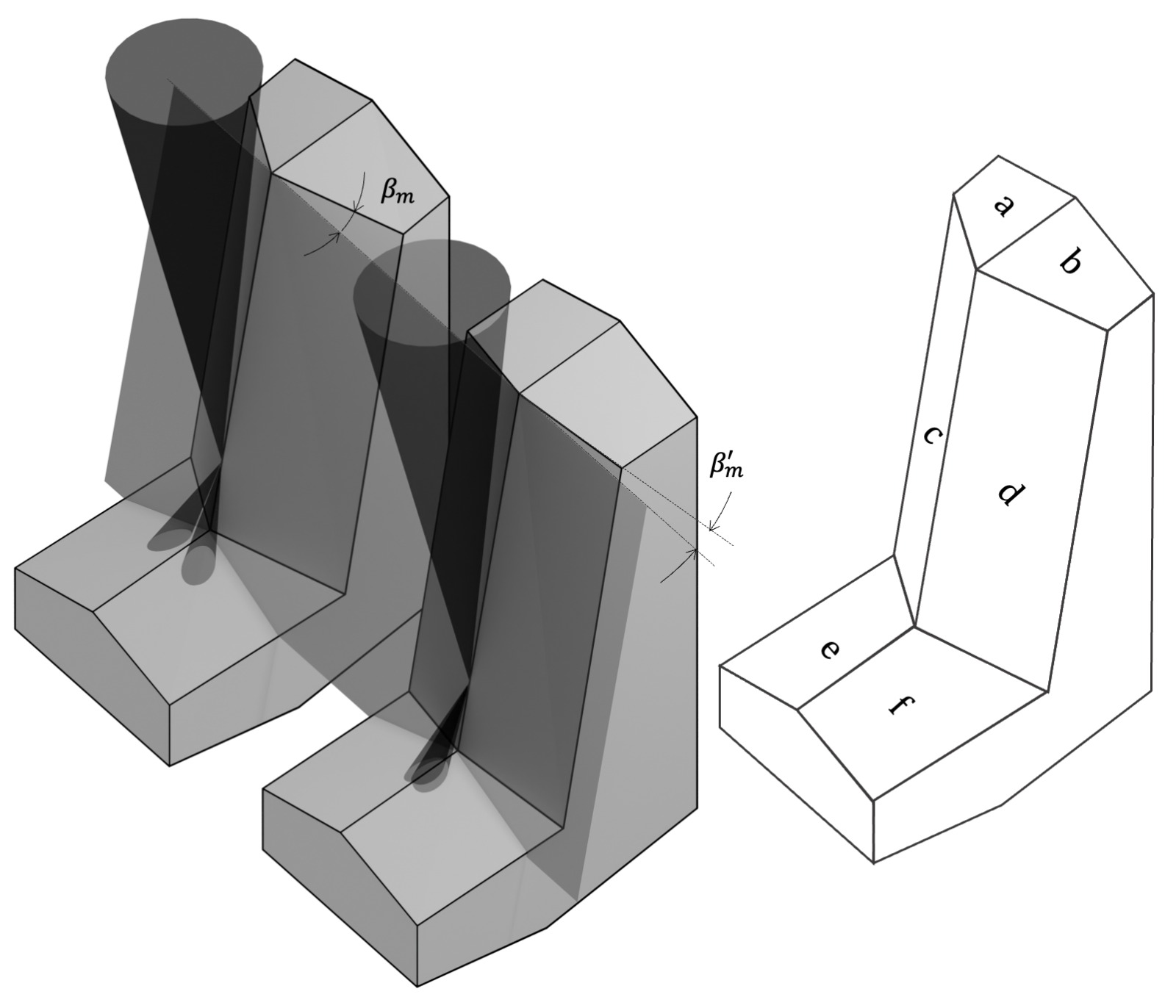}
    \caption{Constraining the prism: the reflecting roof should exhibit a
minimum angle $\beta_m$ in order that the reflection of the beams in the mid part of the LGS source will be properly separated.}
    \label{fig:betam}
\end{figure}

It is to be noted that there are a few additional constraints. For example (Fig.~\ref{fig:betam}) the angle $\beta_m$ should be such that the beams reflected by the faces c and d are separated at the level of the faces e and f.

As $\Delta h$ is continuously changing upon tracking, a further element is needed to define the ingot as an optical element, namely the "height" of such a component, $d_m$. This can be achieved through a number of methods, including suboptimal using of the sensor, an anamorphic or conventional zoom optical relay to match the actual $\Delta h$ with the reimaged length of a fixed ingot, or a variety of ingots with different heights. Solutions noted as 4b and 3 in Table~\ref{tab:variingot} exhibit the practical advantage of not needing to take care of this issue, although they deliberately miss the use of one of the edges of the LGS source to get information on the derivative of the wavefront along the elongation.

\begin{table}[h!]
     \begin{center}
     \begin{tabular}{ c }
     \includegraphics[width=9cm]{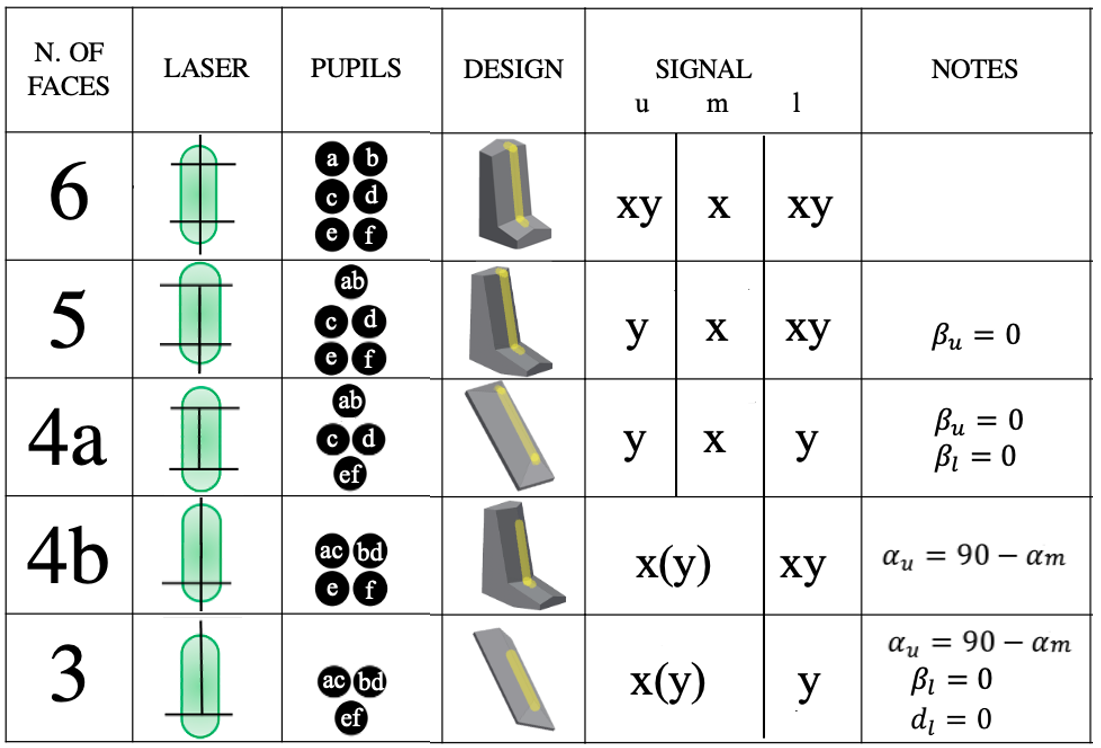}
      \end{tabular}
     \caption{\it{Possible configurations of the ingot prism. Column(1): Number of desirable faces. Column (2)-(3): LGS sections and corresponding pupils. Note that the arrangement of the pupils is purely conceptual, the real position will depend on the considered optic implementation. Column (4): Possible design of the prism. Column (5): Signals that can be measured by considering the upper, medium and lower regions. Column (6): Notes according to Fig.~\ref{fig:3piani}.  } }
     \label{tab:variingot}
      \end{center}
      \end{table}
      
\subsection{An example of dimensioning the ingot prism}
\label{subsec:6pup}

We take the example of an ELT to properly dimension a 6-pupil ingot. The parameters used for the calculation are: the telescope diameter $D=40$ m, laser launcher at $s=21$ m from the telescope axis, focal ratio at the ingot: $F=5$, and a telecentric beam.

The edge between the two reflective faces should be placed along the focal plane of the LGS, which forms an angle with the optical axis given by equation~\ref{eq:finalgeomdem}. This directly determines $\alpha_m\approx6^{\circ}$. We further want the rays reflected by faces c, d to be separated at the level of the faces e, f. This condition is met when the reflected chief rays are separated by at least the cone beam angle:
\begin{equation}
\arccos{\left(\hat{r'_c} \cdot \hat{r'_d}\right)} > 2\arctan{\frac{1}{2F}},
\end{equation}
where $\hat{r'_c}$ and $\hat{r'_d}$ are the ray vectors reflected by the faces c and d respectively, characterized by the angles $\alpha_{m,c} = \alpha_{m,d}$ and $\beta_{m,c}=-\beta_{m,d}$. They can be found using equation~\ref{eq:reflected_ray} or, in the case of small angles, equation~\ref{eq:small_angle_refl}. In our case we get $|\beta_{m}|>36.5^{\circ}$ and, with the small angle approximation, $|\beta_{m}| > 39^{\circ}$.

The angles $\beta_u$ and $\beta_l$ can be easily found from equation~\ref{eq:6pup}, imposing the condition that the distance between the refracted pupils is equal or greater than the diameter. Assuming a refractive index of $n=1.5$, we get $|\beta_{u,l}| > 11.5^{\circ}$.
Consider now the relation for $p_m$ in equation~\ref{eq:6pup}: if the first and second terms in the square brackets have opposite signs and similar amplitudes, they will cancel out and the pupils c and d will overlap. To avoid this, we can choose among two possibilities: 
\begin{enumerate}
    \item choose $\beta_l$ such that the $x$-component of the deviation generated by refraction adds up to the deviation generated by the reflective face. In this case, the lower faces of the ingot (e and f) will form a concave angle. This choice minimizes the angles of incidence and will result in a lower distortion of the pupil images, although the manufacturing of the prism will be likely more challenging and probably the device would be built by gluing together different optical blocks; 
    \item choose $\beta_l$ such that the $x$-component of the deviation generated by refraction compensates the deviation generated by the reflection. In this case, the term inside the square brackets shall be greater than half of the cone beam angle, meaning that $\beta_l$ should be roughly twice the minimum value of $11.5^{\circ}$ calculated above. This option will result in greater angles of incidence and thus higher distortions of the image pupils. The angle between faces e and f will be convex as the one shown in Fig.~\ref{fig:3piani},  leaving the option to build such a device as a single optical element.
\end{enumerate}
We still need to find a proper value for $\alpha_l$, $\alpha_u$ and $\alpha_u'$, all related to the vertical separation of the pupils. Since $\alpha_m$ is fixed by the LGS focal plane, and produces a deviation in the positive $y$ direction according to Fig.~\ref{fig:3piani} (a), it is preferable to choose $\alpha_u$ and $\alpha_u'$ to produce an overall negative deviation. Moreover, to avoid vignetting of the rays refracted by the faces a and b, $\alpha_u$ should also produce a negative or close to zero deviation. 
Also note that $q_m$ and $q_l$ only differ for the term $2\alpha_m f_c $, which is always greater than the pupil diameter when the LGS launcher is outside of the telescope aperture, thus any choice for $\alpha_l$ will keep a proper vertical spacing between the couples c, d and e, f.

Finally, the choice of the angles $\alpha_u, \alpha'_u, \alpha_l$ is somehow arbitrary and deserves a short discussion. Of course, they are to be chosen in order to avoid pupils superimpositions. This is achieved by constraining the distances of the pupil to be larger than their diameter. Furthermore, an overall constant angle can be added to all the angles mentioned above, translating into a simple rigid translation of the pupil images in the detector plane. The detailed optimum choice should be achieved by minimizing the angles of incidence on the ingot faces in order to retain distortions of the pupil images to a minimum. 

As a suboptimal example, choosing $\alpha_l=0$, $\alpha_u'=-\alpha_u$ and imposing the minimum separation between the pupils a,b and e,f, we get $|\alpha_u|>11.5^{\circ}$.

Summarizing, a possible set of proper angles defining a 6-pupils ingot for the ELT case described at the beginning of this subsection, can be individuated as:
\begin{equation}
\begin{gathered}\label{eq:resume}
\begin{split}
\alpha_u  & = 12^{\circ} =- \alpha'_{u}; & &  & \beta_u & = 12^{\circ} =  \beta_l;\\
\alpha_m  & = 6^{\circ};  & & &\beta_m   & = 40^{\circ}; \\
\alpha_l  & = 0^{\circ}.\\
\end{split}
\end{gathered}
\end{equation}

\subsection{3-pupils ingot}
\label{subsec:3pup}
\begin{figure}
    \centering
    \includegraphics[width=7cm]{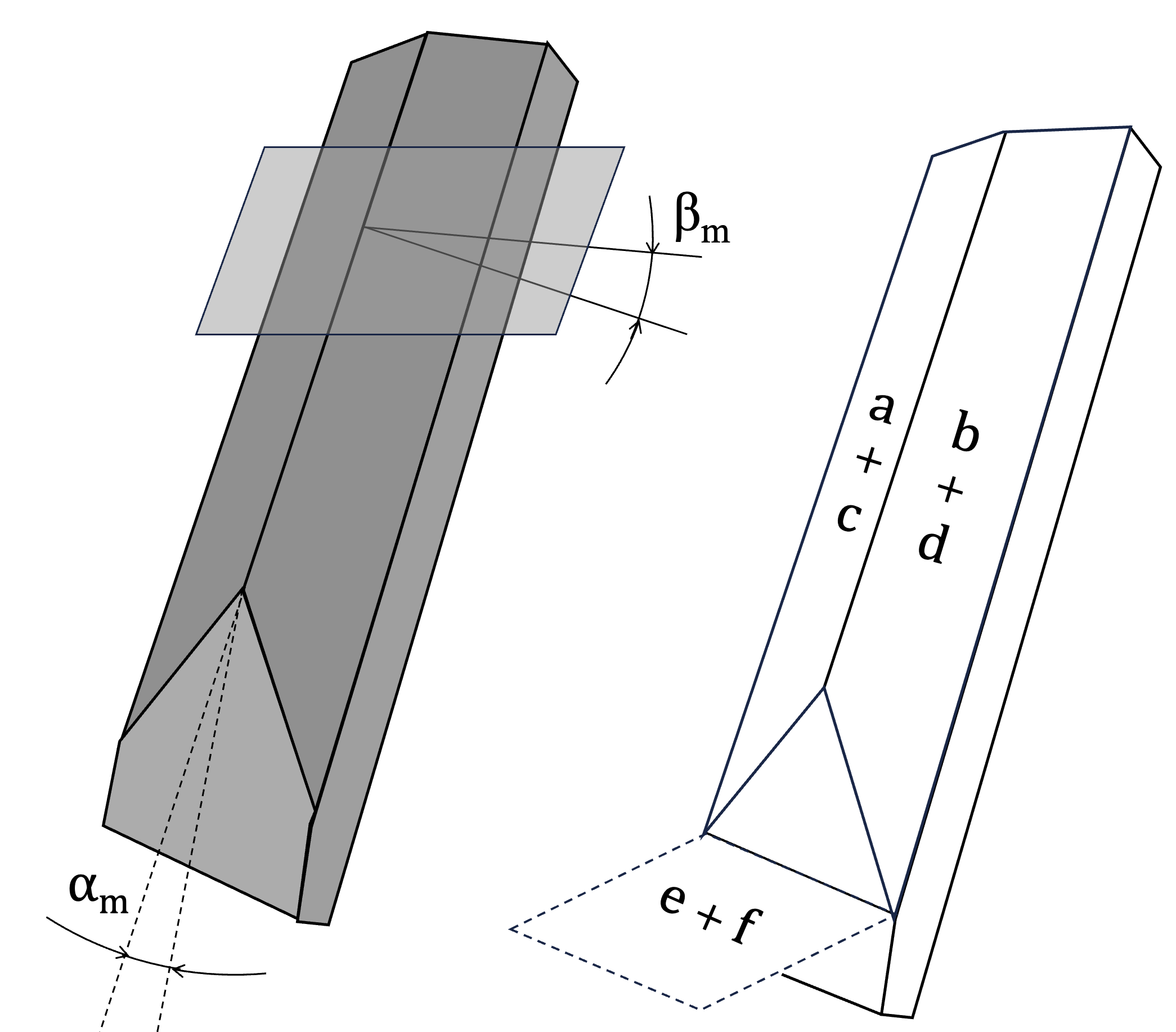}
    \caption{Ingot prism solid shape in the case of 3 pupils. The pupils corresponding to e and f simply go unperturbed. }
    \label{fig:ingot3}
\end{figure}

 In terms of number of reflecting interfaces and corresponding pupils, on the opposite side to the full 6-pupil case described in the previous subsection, it is worth addressing the simplest 3-pupils (case "3" in Table~\ref{tab:variingot}). The practical implementation of this case is particularly advantageous as one of the interfaces (namely the faces e and f) is represented by the light going unperturbed without being reflected or refracted, while the other two interfaces are simple reflections (Fig.~\ref{fig:ingot3}). Along with the "4b" case, this device is inherently compatible with any apparent LGS elongation, at the usual expense of sensing the derivative along $y$ only using one of the edges of the LGS beacon. This edge is chosen as the sharpest end and is conjugated to the lower portion of the Sodium layer, according to the statistics of the layer profiles and using the notion that the resonant molecules "float" over the atmosphere at a somehow well-defined height \citep{Avila1998}.
In practical terms, in such a case, the ingot degenerates into a simple roof, which only should be truncated on one end to avoid the edges introducing some partial vignetting onto the reflected beams. 

It will be characterized by a single angle $\beta_m$ and, using the relationships taken from the Equation~\ref{eq:6pup}, we obtain: 

\begin{equation}
\label{eq:3pup}
\begin{split}
p_m &= f_c [2\beta_m \alpha_m] ;& q_m &= f_c [2\alpha_m];\\
p_l &= 0 ;& q_l &= 0.\\
\end{split}
\end{equation}

Avoiding the pupil superimpositions, and using the same F=5 telescope parameter, as in the previous example, it translates into  $\alpha_m=6^{\circ}$ and $|\beta_{m}|>36.5^{\circ}$.
Although this approach is suboptimal for the measurements of the two derivatives along and orthogonal to the LGS elongation, its benefits are such that these configurations have been the focus of further tests and detailed studies.
Obviously, appropriate choices of the parameters would improve the sensing, reducing the suboptimality to a marginal amount.
A feasibility study for the 3-pupils ingot exploring its implementation on a large telescope and means to explore its sensitivity and resolution has been carried out \citep{Portaluri2022, Portaluri2023}. 

Furthermore, the 3-pupils ingot has been subjected to laboratory testing \citep{DiFilippo2019}, in order to analyze experimentally its sensitivity \citep{2020SPIEKalyan}, by using a simulated LGS whose brightness distribution along the line of propagation can be properly adjusted to mimic the actual measurements \citep{GomesMachado2023}. Low-order quasi-static aberrations measurable by such a WFS have been used to implement a tool to automatically align the ingot to the three-dimensional images of the LGS as collected by a simulated telescope in a real hands-on optical bench \citep{DiFilippo2022}. 
Actual closed-loop experimental simulations on an optical bench have also been carried out with a simplified model of the ingot \citep{Arcidiacono2020}. 

Finally, purely numerical simulation tools have been developed \citep{Viotto2018} to evaluate its performance \citep{Portaluri2019}. This has been accomplished by using a hybrid approach between a pure ray tracing technique and a Fourier analysis \citep{Viotto2019}, and investigating different approximations of the LGS, by adopting several models for the ingot prism, and corresponding measurements of the slopes \citep{2020SPIEElisa}.
 The code includes also the possibility of selecting 3 types of ingot, the 6-, 4- and 3-pupils configuration.

\section{Conclusion}
\label{sec:results}

In this paper, we have formalized the description of a new class of WFSs, nicknamed ingot WFS, which extends the pyramid WFS concept to an elongated LGS. 

We defined several possible optomechanical options, going in a range from 3 to 6 reimaged pupils, and we gave the rationale for computing the proper choice of the geometrical characteristics (angle and size) of the reflecting and refracting surfaces. We fully described the ingot prism as an optical perturbator to be introduced at the reimaged volume, where an LGS is being focused by a large telescope. After that, a common collimator would produce a number of pupil images that can be used in a linearized manner to estimate the derivative of the WF in the two axes.

The optimal use of the detector is one of the advantages that is fully retained with a pyramid-like WFS, and, above all, this is totally independent of the diameter of the telescope aperture. This makes the ingot WFS attractive for ELTs, especially if the lack of large format detectors would lead to the need for truncation of the light, as occurs with the traditional SH WFSs. Some of the several possible implementations and variations are particularly easy to realize, accounting for only a minimum deterioration of the performance, which, of course, should be compared to the capabilities of other conventional WFSs used to sense LGSs. We recall that this new class of WFSs applies only to the cases where the LGS is propagated from outside the telescope aperture and, in fact, it complements an already existing family of WFSs (the z-invariant ones) that were similarly introduced for the case of an LGS being propagated from behind the cage of the secondary mirrors. They both describe a class of pupil plane WFSs aiming specifically to sense the atmospheric turbulence taking into account that the reference light comes from a source extended both on axial and lateral terms and considering the three-dimensional nature of the LGS itself. This is in contrast with most of other approaches where a WFS conceived for a natural reference star is being used in a suboptimal manner, to achieve sensing using only a portion of the light coming from different ranges with respect to the telescope. 

The actual detailed computation of the sensitivity of such new class of WFSs is beyond the scope of this manuscript.
However one can easily note that the ingot approach is the pupil plane version of a SH-like WFS where the spots are reimaged onto a pixel pattern that is aligned, per each subaperture, with the elongation of the LGS itself. 

\begin{figure*}
    \centering
    \includegraphics[width=17.5cm]{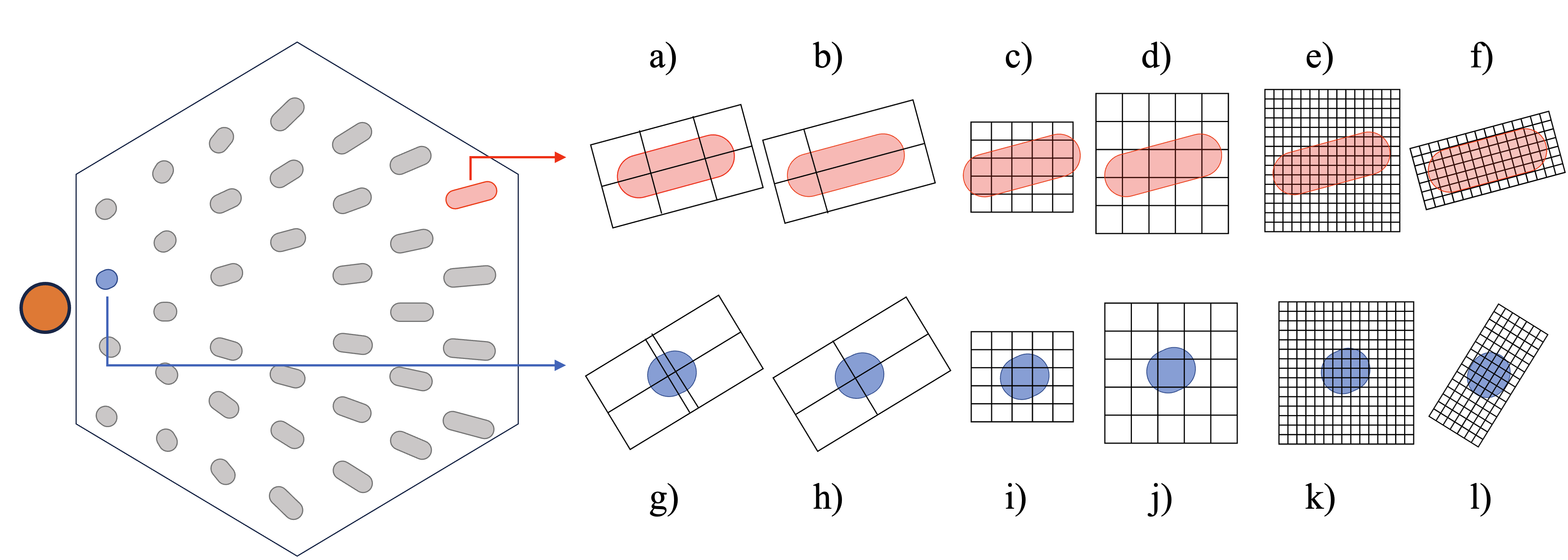}
    \caption{ A didascalic comparison among the ingot and SH WFSs used with an LGS fired aside (propagated from the orange aperture left to the pupil) the hexagonal shaped telescope aperture. As the elongation varies across the entrance pupil, two cases of very mild (blue spots) and extreme (red spots) elongations are taken as representative in the 6-pupils and 3-pupils ingot configurations and for the SH approach. Of course, in the 6-pupils ingot case, a single device is used but its projected elongation on the LGS pattern is inherently adjusted as depicted in cases a) and g) respectively, while cases b) and h) refer to the 3-pupils ingot case. For the SH approach various examples are shown with nominal sampling as in i), which could translate into the need for truncation as depicted in case c) for a limited format detector. This can be overridden with coarse sampling as in cases d) and j) or with a large format detector as in cases e) and k). The use of specifically designed polar detectors, briefly mentioned in the text, is outlined for cases f) and l) where of course the general case of an arbitrary angle between the pixel arrangement and the LGS elongation. It is clear that, as long as light distributions across the elongation is not used, the SH case is suboptimal with respect to the 6-ingot case and that the 3-ingot case is lacking with respect to the former just along the elongated axis $y$.}
    \label{fig:WFScomparison}
\end{figure*}

In order to qualitatively compare the ingot WFS sensitivity with respect of the SH case, one should compare a number of different situations. With reference to the illustration depicted in Fig.~\ref{fig:WFScomparison}, the case of the full 6 faces ingot (this is true also for the case 4b) makes the ultimate use of the LGS light to sense the derivative along the $x$ axis. The 3-pupils ingot would only underperform by the lack of a small portion of the light and it is likely to give comparable results. However, both the cases of LGS truncation, of the undersampling and of the non aligned pixels pattern are clearly expected to underperform. The exceptions are if the whole ensemble of the LGS spots are aligned with the pixel pattern, and the adoption of a sophisticated algorithm, maybe together by a finer sampling of the spot. However, all these options lead to the need of an overall number of pixels of one or two orders of magnitude larger, if we consider an ELT-class telescope apertures. We do not speculate on the consequent additional noise due to this pixel request (for example, the associated increase of Readout Noise). Along the $y$ axis the situation is, of course, radically different. If one plans to take advantage of the (evolving) structure of the light distribution along the LGS, this is clearly unattainable by the ingot concept. However, descarding this approach, the use of the light is almost optimal for the 6-pupils ingot (along with the options  5 and 4a, with ref. to Table~\ref{tab:variingot}) and, in the worst case hampered by a factor two for the other configurations. As all these solutions have proper sense with large aperture and multiple LGSs are to be considered anyway, one should recall that the lack of derivative in one axes for a single LGS is usually constrained by the proper component of other LGS in any of the various MCAO or tomographic schemes.

 The case of polar CCDs developed in the case of an LGS propagated from behind the cage of the obstruction of the telescope is well treated elsewhere \citep{Thomas2008,Adkins2012}. In the case, described in this work, of an LGS fired from outside the pupil plane, the polar detector for a SH WFS would exhibit an asymmetric pattern as the center of the elongation would lie outside from the pupil area. It is reasonable, in fact, to assume that the ingot approach would get similar sensitivity to such an approach with the obvious advantages of a much more compact detector format (and the related consequences in case of a significant read out noise). 

Further numerical and optical bench tests are ongoing while testing on the sky would be an actual breakthrough for such an approach.

\begin{acknowledgements}
The authors would like to thank the Adaptive Optics group at the Laboratoire d'Astrophysique de Marseille for the several, very helpful discussions and to the anonymous Referee for several suggestions that greatly improved the manuscript.
This study has been possible also with the help of ADONI, the Italian National Laboratory for Adaptive Optics and its related financial support.
K.K.R.S. acknowledges support from the INAF Progetto Premiale "Ottica Adattiva Made in Italy per i grandi telescopi del futuro".
E.P. acknowledges the Osservatorio Astronomico di Padova for the hospitality while this paper was in progress.

\end{acknowledgements}

\bibliographystyle{aa} % style aa.bst
\bibliography{ingot_ref}

\begin{thebibliography}{89}
\expandafter\ifx\csname natexlab\endcsname\relax\def\natexlab#1{#1}\fi

\bibitem[{{Adkins}(2012)}]{Adkins2012}
{Adkins}, S.~M. 2012, in Society of Photo-Optical Instrumentation Engineers
  (SPIE) Conference Series, Vol. 8447, Adaptive Optics Systems III, ed. B.~L.
  {Ellerbroek}, E.~{Marchetti}, \& J.-P. {V{\'e}ran}, 84470R

\bibitem[{{Agapito} {et~al.}(2022){Agapito}, {Busoni}, {Carl{\`a}}, {Plantet},
  {Esposito}, \& {Ciliegi}}]{Agapito2022}
{Agapito}, G., {Busoni}, L., {Carl{\`a}}, G., {et~al.} 2022, in Society of
  Photo-Optical Instrumentation Engineers (SPIE) Conference Series, Vol. 12185,
  Adaptive Optics Systems VIII, ed. L.~{Schreiber}, D.~{Schmidt}, \&
  E.~{Vernet}, 121858D

\bibitem[{{Arcidiacono} {et~al.}(2020){Arcidiacono}, {Di Filippo}, {Greggio},
  {Radhakrishnan Santakumari}, {Portaluri}, {Bergomi}, {Viotto}, {Magrin},
  {Ragazzoni}, {Marafatto}, {Dima}, {Farinato}, {Janin-Potiron}, {Fusco},
  {Neichel}, {Fauvarque}, \& {Schatz}}]{Arcidiacono2020}
{Arcidiacono}, C., {Di Filippo}, S., {Greggio}, D., {et~al.} 2020, in Society
  of Photo-Optical Instrumentation Engineers (SPIE) Conference Series, Vol.
  11448, Society of Photo-Optical Instrumentation Engineers (SPIE) Conference
  Series, 1144868

\bibitem[{{Avila} {et~al.}(1998){Avila}, {Vernin}, \& {Cuevas}}]{Avila1998}
{Avila}, R., {Vernin}, J., \& {Cuevas}, S. 1998, \pasp, 110, 1106

\bibitem[{{Beckers}(1988)}]{Beckers1988}
{Beckers}, J.~M. 1988, in European Southern Observatory Conference and Workshop
  Proceedings, Vol.~30, Very Large Telescopes and their Instrumentation, Vol.
  2, 693

\bibitem[{{Beckers}(1993)}]{Beckers1993}
{Beckers}, J.~M. 1993, \araa, 31, 13

\bibitem[{{Bharmal} {et~al.}(2018){Bharmal}, {Myers}, \& {Yang}}]{Bharmal2018}
{Bharmal}, N.~A., {Myers}, R.~M., \& {Yang}, H. 2018, in Society of
  Photo-Optical Instrumentation Engineers (SPIE) Conference Series, Vol. 10703,
  Adaptive Optics Systems VI, ed. L.~M. {Close}, L.~{Schreiber}, \&
  D.~{Schmidt}, 1070362

\bibitem[{{Boyer} \& {Ellerbroek}(2016)}]{Boyer2016}
{Boyer}, C. \& {Ellerbroek}, B. 2016, in Society of Photo-Optical
  Instrumentation Engineers (SPIE) Conference Series, Vol. 9909, Adaptive
  Optics Systems V, ed. E.~{Marchetti}, L.~M. {Close}, \& J.-P. {V{\'e}ran},
  990908

\bibitem[{{Buscher} {et~al.}(2002){Buscher}, {Love}, \& {Myers}}]{Buscher2002}
{Buscher}, D.~F., {Love}, G.~D., \& {Myers}, R.~M. 2002, Optics Letters, 27,
  149

\bibitem[{{Butterley} {et~al.}(2006){Butterley}, {Love}, {Wilson}, {Myers}, \&
  {Morris}}]{SPLASH}
{Butterley}, T., {Love}, G.~D., {Wilson}, R.~W., {Myers}, R.~M., \& {Morris},
  T.~J. 2006, \mnras, 368, 837

\bibitem[{{Calia} {et~al.}(2014){Calia}, {Hackenberg}, {Holzl{\"o}hner},
  {Lewis}, \& {Pfrommer}}]{Calia2014}
{Calia}, D.~B., {Hackenberg}, W., {Holzl{\"o}hner}, R., {Lewis}, S., \&
  {Pfrommer}, T. 2014, Advanced Optical Technologies, 3, 345

\bibitem[{{Clare} {et~al.}(2020){Clare}, {Weddell}, \& {Le Louarn}}]{Clare2020}
{Clare}, R.~M., {Weddell}, S.~J., \& {Le Louarn}, M. 2020, \ao, 59, 6431

\bibitem[{{Di Filippo} {et~al.}(2021){Di Filippo}, {Greggio}, {Bergomi},
  {Radhakrishnan}, {Portaluri}, {Viotto}, {Arcidiacono}, {Magrin}, {Marafatto},
  {Dima}, {Ragazzoni}, {Janin-Potiron}, {Schatz}, {Neichel}, {Fauvarque}, \&
  {Fusco}}]{DiFilippo2019}
{Di Filippo}, S., {Greggio}, D., {Bergomi}, M., {et~al.} 2021, in Adaptive
  Optics for Extremely Large Telescopes 6th Edition (AO4ELT6) Conference
  Series, arXiv:2101.07742

\bibitem[{{Di Filippo} {et~al.}(2022){Di Filippo}, {Greggio}, {Bergomi},
  {Radhakrishnan Santhakumari}, {Portaluri}, {Arcidiacono}, {Viotto},
  {Ragazzoni}, {Dima}, {Marafatto}, {Farinato}, \& {Magrin}}]{DiFilippo2022}
{Di Filippo}, S., {Greggio}, D., {Bergomi}, M., {et~al.} 2022, in Adaptive
  Optics Systems VIII, ed. L.~{Schreiber}, D.~{Schmidt}, \& E.~{Vernet}, Vol.
  12185, 121854V

\bibitem[{{Dicke}(1975)}]{Dicke1975}
{Dicke}, R.~H. 1975, \apj, 198, 605

\bibitem[{{Diolaiti} {et~al.}(2012){Diolaiti}, {Schreiber}, {Foppiani}, \&
  {Lombini}}]{Diolaiti2012}
{Diolaiti}, E., {Schreiber}, L., {Foppiani}, I., \& {Lombini}, M. 2012, in
  Society of Photo-Optical Instrumentation Engineers (SPIE) Conference Series,
  Vol. 8447, Adaptive Optics Systems III, ed. B.~L. {Ellerbroek},
  E.~{Marchetti}, \& J.-P. {V{\'e}ran}, 84471K

\bibitem[{{Diolaiti} {et~al.}(2003){Diolaiti}, {Tozzi}, {Ragazzoni},
  {Ferruzzi}, {Vernet-Viard}, {Esposito}, {Farinato}, {Ghedina}, \&
  {Riccardi}}]{Diolaiti2003}
{Diolaiti}, E., {Tozzi}, A., {Ragazzoni}, R., {et~al.} 2003, in Society of
  Photo-Optical Instrumentation Engineers (SPIE) Conference Series, Vol. 4839,
  Adaptive Optical System Technologies II, ed. P.~L. {Wizinowich} \&
  D.~{Bonaccini}, 299--306

\bibitem[{{D'Orgeville} \& {Fetzer}(2016)}]{DOrgeville2016}
{D'Orgeville}, C. \& {Fetzer}, G.~J. 2016, in Society of Photo-Optical
  Instrumentation Engineers (SPIE) Conference Series, Vol. 9909, Adaptive
  Optics Systems V, ed. E.~{Marchetti}, L.~M. {Close}, \& J.-P. {V{\'e}ran},
  99090R

\bibitem[{{Ellerbroek}(1994)}]{Ellerbroek1994}
{Ellerbroek}, B.~L. 1994, Journal of the Optical Society of America A, 11, 783

\bibitem[{{Esposito} \& {Busoni}(2008)}]{Esposito2008}
{Esposito}, S. \& {Busoni}, L. 2008, in Society of Photo-Optical
  Instrumentation Engineers (SPIE) Conference Series, Vol. 7015, Adaptive
  Optics Systems, ed. N.~{Hubin}, C.~E. {Max}, \& P.~L. {Wizinowich}, 70151P

\bibitem[{{Esposito} {et~al.}(2000){Esposito}, {Ragazzoni}, {Riccardi},
  {O'Sullivan}, {Ageorges}, {Redfern}, \& {Davies}}]{Esposito2000}
{Esposito}, S., {Ragazzoni}, R., {Riccardi}, A., {et~al.} 2000, Experimental
  Astronomy, 10, 135

\bibitem[{{Esposito} {et~al.}(2010){Esposito}, {Riccardi}, {Fini}, {Puglisi},
  {Pinna}, {Xompero}, {Briguglio}, {Quir{\'o}s-Pacheco}, {Stefanini}, {Guerra},
  {Busoni}, {Tozzi}, {Pieralli}, {Agapito}, {Brusa-Zappellini}, {Demers},
  {Brynnel}, {Arcidiacono}, \& {Salinari}}]{Esposito2010}
{Esposito}, S., {Riccardi}, A., {Fini}, L., {et~al.} 2010, in Society of
  Photo-Optical Instrumentation Engineers (SPIE) Conference Series, Vol. 7736,
  Adaptive Optics Systems II, ed. B.~L. {Ellerbroek}, M.~{Hart}, N.~{Hubin}, \&
  P.~L. {Wizinowich}, 773609

\bibitem[{{Esposito} {et~al.}(2003){Esposito}, {Tozzi}, {Ferruzzi},
  {Carbillet}, {Riccardi}, {Fini}, {V{\'e}rinaud}, {Accardo}, {Brusa},
  {Gallieni}, {Biasi}, {Baffa}, {Biliotti}, {Foppiani}, {Puglisi}, {Ragazzoni},
  {Ranfagni}, {Stefanini}, {Salinari}, {Seifert}, \& {Storm}}]{Esposito2003}
{Esposito}, S., {Tozzi}, A., {Ferruzzi}, D., {et~al.} 2003, in Society of
  Photo-Optical Instrumentation Engineers (SPIE) Conference Series, Vol. 4839,
  Adaptive Optical System Technologies II, ed. P.~L. {Wizinowich} \&
  D.~{Bonaccini}, 164--173

\bibitem[{{Foy} \& {Labeyrie}(1985)}]{Foy1985}
{Foy}, R. \& {Labeyrie}, A. 1985, \aap, 152, L29

\bibitem[{{Foy} {et~al.}(1995){Foy}, {Migus}, {Biraben}, {Grynberg},
  {McCullough}, \& {Tallon}}]{Foy1995}
{Foy}, R., {Migus}, A., {Biraben}, F., {et~al.} 1995, \aaps, 111, 569

\bibitem[{{Fugate} {et~al.}(1994){Fugate}, {Ellerbroek}, {Higgins}, {Jelonek},
  {Lange}, {Slavin}, {Wild}, {Winker}, {Wynia}, {Spinhirne}, {Boeke}, {Ruane},
  {Moroney}, {Oliker}, {Swindle}, \& {Cleis}}]{Fugate1994}
{Fugate}, R.~Q., {Ellerbroek}, B.~L., {Higgins}, C.~H., {et~al.} 1994, Journal
  of the Optical Society of America A, 11, 310

\bibitem[{{Genzel} {et~al.}(2003){Genzel}, {Sch{\"o}del}, {Ott}, {Eisenhauer},
  {Hofmann}, {Lehnert}, {Eckart}, {Alexander}, {Sternberg}, {Lenzen},
  {Cl{\'e}net}, {Lacombe}, {Rouan}, {Renzini}, \&
  {Tacconi-Garman}}]{Genzel2003}
{Genzel}, R., {Sch{\"o}del}, R., {Ott}, T., {et~al.} 2003, \apj, 594, 812

\bibitem[{{Ghez} {et~al.}(2008){Ghez}, {Salim}, {Weinberg}, {Lu}, {Do}, {Dunn},
  {Matthews}, {Morris}, {Yelda}, {Becklin}, {Kremenek}, {Milosavljevic}, \&
  {Naiman}}]{Ghez2008}
{Ghez}, A.~M., {Salim}, S., {Weinberg}, N.~N., {et~al.} 2008, \apj, 689, 1044

\bibitem[{{Gilmozzi} \& {Spyromilio}(2007)}]{Gilmozzi2007}
{Gilmozzi}, R. \& {Spyromilio}, J. 2007, The Messenger, 127

\bibitem[{{Gomes Machado} {et~al.}(2023){Gomes Machado}, {Di Filippo},
  {Bergomi}, {Kumar Radhakrishnan Santhakumari}, {Greggio}, {Portaluri},
  {Arcidiacono}, {Viotto}, {Ragazzoni}, {Dima}, {Marafatto}, {Farinato}, \&
  {Magrin}}]{GomesMachado2023}
{Gomes Machado}, T.~S., {Di Filippo}, S., {Bergomi}, M., {et~al.} 2023, in
  Adaptive Optics for Extremely Large Telescopes 7th Edition (AO4ELT7)
  Conference Series

\bibitem[{{Gratadour} {et~al.}(2010){Gratadour}, {Gendron}, {Rousset}, \&
  {Rigaut}}]{Gratadour2010}
{Gratadour}, D., {Gendron}, E., {Rousset}, G., \& {Rigaut}, F. 2010, in
  Adaptative Optics for Extremely Large Telescopes, 04005

\bibitem[{{Guyon}(2018)}]{Guyon2018}
{Guyon}, O. 2018, \araa, 56, 315

\bibitem[{{Hardy}(1998)}]{Hardy1998}
{Hardy}, J.~W. 1998, {Adaptive Optics for Astronomical Telescopes}

\bibitem[{{Herbst} {et~al.}(2018){Herbst}, {Santhakumari}, {Klettke},
  {Arcidiacono}, {Bergomi}, {Bertram}, {Berwein}, {Bizenberger}, {Briegel},
  {Farinato}, {Marafatto}, {Mathar}, {McGurk}, {Ragazzoni}, \&
  {Viotto}}]{Herbst2018}
{Herbst}, T.~M., {Santhakumari}, K. K.~R., {Klettke}, M., {et~al.} 2018, in
  Society of Photo-Optical Instrumentation Engineers (SPIE) Conference Series,
  Vol. 10703, Adaptive Optics Systems VI, ed. L.~M. {Close}, L.~{Schreiber}, \&
  D.~{Schmidt}, 107030B

\bibitem[{{Herrmann}(1992)}]{Herrmann1992}
{Herrmann}, J. 1992, Journal of the Optical Society of America A, 9, 2257

\bibitem[{{Horwitz}(1994)}]{Horwitz1994}
{Horwitz}, B.~A. 1994, in Society of Photo-Optical Instrumentation Engineers
  (SPIE) Conference Series, Vol. 2201, Adaptive Optics in Astronomy, ed. M.~A.
  {Ealey} \& F.~{Merkle}, 496--501

\bibitem[{{Johns}(2008)}]{Johns2008}
{Johns}, M. 2008, in \procspie, Vol. 6986, Extremely Large Telescopes: Which
  Wavelengths? Retirement Symposium for Arne Ardeberg, 698603

\bibitem[{{Lombini} {et~al.}(2022){Lombini}, {Schreiber}, {Diolaiti}, \&
  {Cortecchia}}]{Lombini2022}
{Lombini}, M., {Schreiber}, L., {Diolaiti}, E., \& {Cortecchia}, F. 2022,
  \mnras, 510, 3876

\bibitem[{{Ma} \& {Wang}(2016)}]{Ma2016}
{Ma}, X. \& {Wang}, J. 2016, Optik, 127, 2688

\bibitem[{{Macintosh}(2001)}]{Macintosh2001}
{Macintosh}, B. 2001, in American Astronomical Society Meeting Abstracts, Vol.
  198, American Astronomical Society Meeting Abstracts \#198, 83.04

\bibitem[{{Marchetti} {et~al.}(2008){Marchetti}, {Brast}, {Delabre},
  {Donaldson}, {Fedrigo}, {Frank}, {Hubin}, {Kolb}, {Lizon}, {Marchesi},
  {Oberti}, {Reiss}, {Soenke}, {Tordo}, {Baruffolo}, {Bagnara}, {Amorim}, \&
  {Lima}}]{Marchetti2008}
{Marchetti}, E., {Brast}, R., {Delabre}, B., {et~al.} 2008, in Society of
  Photo-Optical Instrumentation Engineers (SPIE) Conference Series, Vol. 7015,
  Adaptive Optics Systems, ed. N.~{Hubin}, C.~E. {Max}, \& P.~L. {Wizinowich},
  70150F

\bibitem[{Marchetti {et~al.}(2003)Marchetti, Hubin, Fedrigo, Brynnel, Delabre,
  Donaldson, Franza, Conan, Le~Louarn, Cavadore, Balestra, Baade, Lizon,
  Gilmozzi, Monnet, Ragazzoni, Arcidiacono, Baruffolo, Diolaiti, Farinato,
  Vernet-Viard, Butler, Hippler, \& Amorin}]{marchetti_mad_2003}
Marchetti, E., Hubin, N.~N., Fedrigo, E., {et~al.} 2003, 4839, 317, conference
  Name: Adaptive Optical System Technologies II ADS Bibcode:
  2003SPIE.4839..317M

\bibitem[{{Mayer}(1994)}]{Scheim1994}
{Mayer}, H. 1994, Ophthalmic Res, 26, 3

\bibitem[{{Neichel} {et~al.}(2014){Neichel}, {Rigaut}, {Vidal}, {van Dam},
  {Garrel}, {Carrasco}, {Pessev}, {Winge}, {Boccas}, {D'Orgeville},
  {Arriagada}, {Serio}, {Fesquet}, {Rambold}, {L{\"u}hrs}, {Moreno},
  {Gausachs}, {Galvez}, {Montes}, {Vucina}, {Marin}, {Urrutia}, {Lopez},
  {Diggs}, {Marchant}, {Ebbers}, {Trujillo}, {Bec}, {Trancho}, {McGregor},
  {Young}, {Colazo}, \& {Edwards}}]{Neichel2014}
{Neichel}, B., {Rigaut}, F., {Vidal}, F., {et~al.} 2014, \mnras, 440, 1002

\bibitem[{{Pilkington}(1987)}]{Pilkington1987}
{Pilkington}, J.~D.~H. 1987, \nat, 330, 116

\bibitem[{{Portaluri} {et~al.}(2022){Portaluri}, {Di Filippo}, {Viotto},
  {Ragazzoni}, {Arcidiacono}, {Greggio}, {Radhakrishnan Santhakumari},
  {Bergomi}, {Marafatto}, {Dima}, {Farinato}, {Magrin}, {Di Rico}, {Centrone},
  \& {Bonaccini}}]{Portaluri2022}
{Portaluri}, E., {Di Filippo}, S., {Viotto}, V., {et~al.} 2022, in Society of
  Photo-Optical Instrumentation Engineers (SPIE) Conference Series, Vol. 12185,
  Adaptive Optics Systems VIII, ed. L.~{Schreiber}, D.~{Schmidt}, \&
  E.~{Vernet}, 121851K

\bibitem[{{Portaluri} {et~al.}(2023){Portaluri}, {Radhakrishnan Santhakumari},
  {Ragazzoni}, {Greggio}, {Arcidiacono}, {Bergomi}, {Di Filippo}, {Dima},
  {Farinato}, {Gomes Machado}, {Magrin}, \& {Viotto}}]{Portaluri2023}
{Portaluri}, E., {Radhakrishnan Santhakumari}, K., {Ragazzoni}, R., {et~al.}
  2023, in Adaptive Optics for Extremely Large Telescopes 7th Edition (AO4ELT7)
  Conference Series

\bibitem[{{Portaluri} {et~al.}(2020{\natexlab{a}}){Portaluri}, {Viotto},
  {Ragazzoni}, {Arcidiacono}, {Bergomi}, {Greggio}, {Radhakrishnan}, {di
  Filippo}, {Marafatto}, {Dima}, {Biondi}, {Farinato}, \&
  {Magrin}}]{Portaluri2019}
{Portaluri}, E., {Viotto}, V., {Ragazzoni}, R., {et~al.} 2020{\natexlab{a}}, in
  Adaptive Optics for Extremely Large Telescopes 6th Edition (AO4ELT6)
  Conference Series, arXiv:2012.09514

\bibitem[{{Portaluri} {et~al.}(2020{\natexlab{b}}){Portaluri}, {Viotto},
  {Ragazzoni}, {Arcidiacono}, {Greggio}, {Radhakrishnan Santhakumari},
  {Bergomi}, {Di Filippo}, {Farinato}, \& {Magrin}}]{2020SPIEElisa}
{Portaluri}, E., {Viotto}, V., {Ragazzoni}, R., {et~al.} 2020{\natexlab{b}}, in
  Society of Photo-Optical Instrumentation Engineers (SPIE) Conference Series,
  Vol. 11448, 114483I

\bibitem[{{Portaluri} {et~al.}(2017){Portaluri}, {Viotto}, {Ragazzoni},
  {Gullieuszik}, {Bergomi}, {Greggio}, {Biondi}, {Dima}, {Magrin}, \&
  {Farinato}}]{Portaluri2017}
{Portaluri}, E., {Viotto}, V., {Ragazzoni}, R., {et~al.} 2017, \mnras, 466,
  3569

\bibitem[{{Radhakrishnan Santhakumari} {et~al.}(2020){Radhakrishnan
  Santhakumari}, {Greggio}, {Bergomi}, {Di Filippo}, {Viotto}, {Portaluri},
  {Arcidiacono}, {Dima}, {Lessio}, {Marafatto}, {Furieri}, {Bonora}, \&
  {Ragazzoni}}]{2020SPIEKalyan}
{Radhakrishnan Santhakumari}, K.~K., {Greggio}, D., {Bergomi}, M., {et~al.}
  2020, in Society of Photo-Optical Instrumentation Engineers (SPIE) Conference
  Series, Vol. 11448, Society of Photo-Optical Instrumentation Engineers (SPIE)
  Conference Series, 1144860

\bibitem[{{Ragazzoni}(1996{\natexlab{a}})}]{Ragazzoni1996}
{Ragazzoni}, R. 1996{\natexlab{a}}, \apjl, 465, L73

\bibitem[{{Ragazzoni}(1996{\natexlab{b}})}]{Ragazzoni1996b}
{Ragazzoni}, R. 1996{\natexlab{b}}, Journal of Modern Optics, 43, 289

\bibitem[{{Ragazzoni}(1997)}]{Ragazzoni1997}
{Ragazzoni}, R. 1997, \aap, 319, L9

\bibitem[{{Ragazzoni}(2000)}]{Ragazzoni2000nato}
{Ragazzoni}, R. 2000, in NATO Advanced Study Institute (ASI) Series C, Vol.
  551, Laser Guide Star Adaptive Optics for Astronomy, ed. N.~{Ageorges} \&
  C.~{Dainty}, 125

\bibitem[{{Ragazzoni}(2001)}]{Ragazzoni2001b}
{Ragazzoni}, R. 2001, in Science with the Large Binocular Telescope, ed.
  T.~{Herbst}, 13

\bibitem[{{Ragazzoni}(2014)}]{Ragazzoni2014}
{Ragazzoni}, R. 2014, in Society of Photo-Optical Instrumentation Engineers
  (SPIE) Conference Series, Vol. 9148, Adaptive Optics Systems IV, ed.
  E.~{Marchetti}, L.~M. {Close}, \& J.-P. {Vran}, 914811

\bibitem[{{Ragazzoni} {et~al.}(2002){Ragazzoni}, {Diolaiti}, {Farinato},
  {Fedrigo}, {Marchetti}, {Tordi}, \& {Kirkman}}]{Ragazzoni2002}
{Ragazzoni}, R., {Diolaiti}, E., {Farinato}, J., {et~al.} 2002, \aap, 396, 731

\bibitem[{{Ragazzoni} \& {Farinato}(1999)}]{Ragazzoni1999}
{Ragazzoni}, R. \& {Farinato}, J. 1999, \aap, 350, L23

\bibitem[{{Ragazzoni} {et~al.}(2000{\natexlab{a}}){Ragazzoni}, {Farinato}, \&
  {Marchetti}}]{Ragazzoni2000}
{Ragazzoni}, R., {Farinato}, J., \& {Marchetti}, E. 2000{\natexlab{a}}, in
  Society of Photo-Optical Instrumentation Engineers (SPIE) Conference Series,
  Vol. 4007, Adaptive Optical Systems Technology, ed. P.~L. {Wizinowich},
  1076--1087

\bibitem[{{Ragazzoni} {et~al.}(2000{\natexlab{b}}){Ragazzoni}, {Giallongo},
  {Pasian}, {Pedichini}, {Fontana}, {Marconi}, {Speziali}, {Turatto},
  {Danziger}, {Cremonese}, {Smareglia}, {Gallieni}, {Anaclerio}, \&
  {Lazzarini}}]{Ragazzoni2000b}
{Ragazzoni}, R., {Giallongo}, E., {Pasian}, F., {et~al.} 2000{\natexlab{b}}, in
  Society of Photo-Optical Instrumentation Engineers (SPIE) Conference Series,
  Vol. 4008, Optical and IR Telescope Instrumentation and Detectors, ed.
  M.~{Iye} \& A.~F. {Moorwood}, 439--446

\bibitem[{{Ragazzoni} {et~al.}(2018){Ragazzoni}, {Greggio}, {Viotto}, {Di
  Filippo}, {Dima}, {Farinato}, {Bergomi}, {Portaluri}, {Magrin}, {Marafatto},
  {Biondi}, {Carolo}, {Chinellato}, {Umbriaco}, \& {Vassallo}}]{Ragazzoni2018}
{Ragazzoni}, R., {Greggio}, D., {Viotto}, V., {et~al.} 2018, in Society of
  Photo-Optical Instrumentation Engineers (SPIE) Conference Series, Vol. 10703,
  Adaptive Optics Systems VI, ed. L.~M. {Close}, L.~{Schreiber}, \&
  D.~{Schmidt}, 107033Y

\bibitem[{{Ragazzoni} {et~al.}(2006){Ragazzoni}, {Kellner}, {Gaessler},
  {Diolaiti}, \& {Farinato}}]{Ragazzoni2006}
{Ragazzoni}, R., {Kellner}, S., {Gaessler}, W., {Diolaiti}, E., \& {Farinato},
  J. 2006, \mnras, 368, 1796

\bibitem[{{Ragazzoni} {et~al.}(2017){Ragazzoni}, {Portaluri}, {Viotto}, {Dima},
  {Bergomi}, {Biondi}, {Farinato}, {Carolo}, {Chinellato}, {Greggio},
  {Gullieuszik}, {Magrin}, {Marafatto}, \& {Vassallo}}]{Ragazzoni2017}
{Ragazzoni}, R., {Portaluri}, E., {Viotto}, V., {et~al.} 2017, in Adaptive
  Optics for Extremely Large Telescopes 5th Edition (AO4ELT5) Conference Series

\bibitem[{{Ragazzoni} {et~al.}(2001){Ragazzoni}, {Tordi}, {Diolaiti}, \&
  {Kirkman}}]{Ragazzoni2001}
{Ragazzoni}, R., {Tordi}, M., {Diolaiti}, E., \& {Kirkman}, D. 2001, \mnras,
  327, 949

\bibitem[{{Ragazzoni} {et~al.}(2013){Ragazzoni}, {Viotto}, {Magrin}, {Bergomi},
  {Dima}, {Farinato}, {Greggio}, \& {Marafatto}}]{Ragazzoni2013}
{Ragazzoni}, R., {Viotto}, V., {Magrin}, D., {et~al.} 2013, in Adaptive Optics
  for Extremely Large Telescopes 3th Edition (AO4ELT3) Conference Series, ed.
  S.~{Esposito} \& L.~{Fini}, 33

\bibitem[{{Ragazzoni} {et~al.}(2019){Ragazzoni}, Viotto, Portaluri, Bergomi,
  Greggio, {Di Filippo}, Radhakrishnan, Umbriaco, Dima, Magrin, Farinato,
  Marafatto, Arcidiacono, \& Biondi}]{Ragazzoni2019}
{Ragazzoni}, R., Viotto, V., Portaluri, E., {et~al.} 2019, in Adaptive Optics
  for Extremely Large Telescopes 6th Edition (AO4ELT6) Conference Series

\bibitem[{{Rigaut}(2002)}]{Rigaut2002}
{Rigaut}, F. 2002, in European Southern Observatory Conference and Workshop
  Proceedings, Vol.~58, European Southern Observatory Conference and Workshop
  Proceedings, 11

\bibitem[{{Rigaut} \& {Gendron}(1992)}]{Rigaut1992}
{Rigaut}, F. \& {Gendron}, E. 1992, \aap, 261, 677

\bibitem[{{Rigaut} \& {Neichel}(2020)}]{Rigaut2020}
{Rigaut}, F. \& {Neichel}, B. 2020, arXiv e-prints, arXiv:2003.03097

\bibitem[{{Rigaut} {et~al.}(2014){Rigaut}, {Neichel}, {Boccas}, {D'Orgeville},
  {Vidal}, {van Dam}, {Arriagada}, {Fesquet}, {Galvez}, {Gausachs}, {Cavedoni},
  {Ebbers}, {Karewicz}, {James}, {L{\"u}hrs}, {Montes}, {Perez}, {Rambold},
  {Rojas}, {Walker}, {Bec}, {Trancho}, {Sheehan}, {Irarrazaval}, {Boyer},
  {Ellerbroek}, {Flicker}, {Gratadour}, {Garcia-Rissmann}, \&
  {Daruich}}]{Rigaut2014}
{Rigaut}, F., {Neichel}, B., {Boccas}, M., {et~al.} 2014, \mnras, 437, 2361

\bibitem[{{Rodr{\'\i}guez-Ramos} {et~al.}(2008){Rodr{\'\i}guez-Ramos},
  {Femen{\'\i}a Castell{\'a}}, {P{\'e}rez Nava}, \& {Fumero}}]{Ramos2008}
{Rodr{\'\i}guez-Ramos}, J.~M., {Femen{\'\i}a Castell{\'a}}, B., {P{\'e}rez
  Nava}, F., \& {Fumero}, S. 2008, in Society of Photo-Optical Instrumentation
  Engineers (SPIE) Conference Series, Vol. 7015, Adaptive Optics Systems, ed.
  N.~{Hubin}, C.~E. {Max}, \& P.~L. {Wizinowich}, 70155Q

\bibitem[{Scheimpflug(1904)}]{ScheimpflugBrevetto}
Scheimpflug, T. 1904, Improved method and apparatus for the systematic
  alteration or distortion of plane pictures and images by means of lenses and
  mirrors for photography and for other purposes

\bibitem[{{Schreiber} {et~al.}(2014){Schreiber}, {Diolaiti}, {Arcidiacono},
  {Pfrommer}, {Holzl{\"o}hner}, {Lombini}, \& {Hickson}}]{Schreiber2014}
{Schreiber}, L., {Diolaiti}, E., {Arcidiacono}, C., {et~al.} 2014, in Society
  of Photo-Optical Instrumentation Engineers (SPIE) Conference Series, Vol.
  9148, Adaptive Optics Systems IV, ed. E.~{Marchetti}, L.~M. {Close}, \& J.-P.
  {Vran}, 91486Q

\bibitem[{{Strehl}(1895)}]{Strehl1895}
{Strehl}, K. 1895, Zeitschrift für Instrumentenkunde, 15, 362

\bibitem[{{Szeto} {et~al.}(2008){Szeto}, {Roberts}, {Gedig}, {Austin},
  {Lagally}, {Patrick}, {Tsang}, {MacMynowski}, {Sirota}, {Stepp}, \&
  {Thompson}}]{Szeto2008}
{Szeto}, K., {Roberts}, S., {Gedig}, M., {et~al.} 2008, in Society of
  Photo-Optical Instrumentation Engineers (SPIE) Conference Series, Vol. 7012,
  Ground-based and Airborne Telescopes II, ed. L.~M. {Stepp} \& R.~{Gilmozzi},
  70122G

\bibitem[{{Thomas} {et~al.}(2008){Thomas}, {Adkins}, {Gavel}, {Fusco}, \&
  {Michau}}]{Thomas2008}
{Thomas}, S.~J., {Adkins}, S., {Gavel}, D., {Fusco}, T., \& {Michau}, V. 2008,
  \mnras, 387, 173

\bibitem[{{Thompson} \& {Gardner}(1987)}]{Thompson1987}
{Thompson}, L.~A. \& {Gardner}, C.~S. 1987, \nat, 328, 229

\bibitem[{{Tokovinin}(2004)}]{Tokovinin2004}
{Tokovinin}, A. 2004, \pasp, 116, 941

\bibitem[{{Tr{\"a}ger}(2012)}]{2012TextbookLaser}
{Tr{\"a}ger}, F. 2012, {Springer Handbook of Lasers and Optics}, 69

\bibitem[{{Tyson}(1991)}]{Tyson1991}
{Tyson}, R.~K. 1991, {Principles of adaptive optics}

\bibitem[{{Vieira} {et~al.}(2018){Vieira}, {Rodrigues Pipa}, \&
  {Mello}}]{Vieira2018}
{Vieira}, L. E.~L., {Rodrigues Pipa}, D., \& {Mello}, A. J. T.~S. 2018, in
  Society of Photo-Optical Instrumentation Engineers (SPIE) Conference Series,
  Vol. 10772, Unconventional and Indirect Imaging, Image Reconstruction, and
  Wavefront Sensing 2018, ed. J.~J. {Dolne} \& P.~J. {Bones}, 107720B

\bibitem[{{Viotto} {et~al.}(2015){Viotto}, {Bergomi}, {Portaluri}, {Dima},
  {Farinato}, {Greggio}, {Magrin}, \& {Ragazzoni}}]{Viotto2015}
{Viotto}, V., {Bergomi}, M., {Portaluri}, E., {et~al.} 2015, in Adaptive Optics
  for Extremely Large Telescopes 4th Edition (AO4ELT4) Conference Series, E34

\bibitem[{Viotto {et~al.}(2019)Viotto, Portaluri, Arcidiacono, Bergomi, {Di
  Filippo}, Greggio, Radhakrishnan, Dima, Farinato, Magrin, Marafatto, \&
  Ragazzoni}]{Viotto2019}
Viotto, V., Portaluri, E., Arcidiacono, C., {et~al.} 2019, in Adaptive Optics
  for Extremely Large Telescopes 6th Edition (AO4ELT6) Conference Series

\bibitem[{{Viotto} {et~al.}(2018){Viotto}, {Portaluri}, {Arcidiacono},
  {Ragazzoni}, {Bergomi}, {Di Filippo}, {Dima}, {Farinato}, {Greggio},
  {Magrin}, \& {Marafatto}}]{Viotto2018}
{Viotto}, V., {Portaluri}, E., {Arcidiacono}, C., {et~al.} 2018, in Society of
  Photo-Optical Instrumentation Engineers (SPIE) Conference Series, Vol. 10703,
  Adaptive Optics Systems VI, ed. L.~M. {Close}, L.~{Schreiber}, \&
  D.~{Schmidt}, 107030V

\bibitem[{{Wilson} {et~al.}(1997){Wilson}, {Lesh}, {Araki}, \&
  {Arimoto}}]{Wilson1997}
{Wilson}, K.~E., {Lesh}, J.~R., {Araki}, K., \& {Arimoto}, Y. 1997, in Society
  of Photo-Optical Instrumentation Engineers (SPIE) Conference Series, Vol.
  2990, Free-Space Laser Communication Technologies IX, ed. G.~S. {Mecherle},
  23--30

\bibitem[{{Yang} {et~al.}(2019){Yang}, {Bharmal}, {Myers}, \&
  {Younger}}]{Yang2019}
{Yang}, H., {Bharmal}, N., {Myers}, R., \& {Younger}, E. 2019, Journal of
  Astronomical Telescopes, Instruments, and Systems, 5, 029002

\bibitem[{{Yang} {et~al.}(2018){Yang}, {Bharmal}, \& {Myers}}]{Yang2018}
{Yang}, H., {Bharmal}, N.~A., \& {Myers}, R.~M. 2018, \mnras, 477, 4443

\bibitem[{{Zhang} {et~al.}(2021){Zhang}, {Morris}, {Bharmal}, \&
  {Liang}}]{Zhang2021}
{Zhang}, Z., {Morris}, T., {Bharmal}, N., \& {Liang}, Y. 2021, \ao, 60, 4208

\end{thebibliography}

\end{document}